\documentclass{article} \usepackage{amssymb,amsmath}
\usepackage[numbers]{natbib}
\usepackage{graphicx}
\usepackage{subfigure}
\usepackage{multirow,array}
\usepackage{epsfig}
\usepackage{threeparttable}
\usepackage{changes}
\usepackage{authblk}

\title{Verification and comparison of four numerical
schemes for a 1D viscoelastic blood flow model}
\author[1]{Xiaofei Wang}
\author[1]{Jose-Maria Fullana}
\author[2]{Pierre-Yves Lagr\'ee}
\affil[1]{Sorbonne Universit\'es, UPMC Univ Paris 6, UMR 7190, Institut Jean le Rond $\partial$'Alembert}
\affil[2]{CNRS, UMR 7190, Institut Jean le Rond $\partial$'Alembert}

\date{\today}
\begin{document}
\maketitle

\begin{abstract}
A reliable and fast numerical scheme is crucial for the 1D simulation of blood flow in compliant vessels. 
In this paper, a 1D blood flow model is incorporated with a Kelvin-Voigt viscoelastic arterial wall.
This leads to a nonlinear hyperbolic-parabolic system, which is then solved with four numerical schemes, 
namely: MacCormack, Taylor-Galerkin, MUSCL (monotonic upwind scheme for conservation law)
and local discontinuous Galerkin.
The numerical schemes are tested on a single vessel, a simple bifurcation and a network with 55 arteries.
The numerical solutions are checked favorably against analytical,
semi-analytical solutions or clinical observations.  
Among the numerical schemes, comparisons are made in four important aspects:
accuracy, ability to capture shock-like phenomena, computational speed
and implementation complexity. 
The suitable conditions for the application of each scheme are discussed.
\end{abstract}
\let\thefootnote\relax\footnote{This is an Accepted Manuscript of an article published by Taylor \& Francis Group in Computer Methods in Biomechanics and Biomedical Engineering on 22 Aug 2014, available online: http://www.tandfonline.com/doi/full/10.1080/10255842.2014.948428.}


\section{Introduction}

Simulating the blood flow in compliant vessels is of great clinical relevance and is also a challenging problem. 
Many 3D simulations of this fluid-structure interaction (FSI) are presented in literature~\cite{bertoglio2012sequential,
crosetto2011fluid,di2001fluid,gerbeau2005fluid,li2005blood,perktold1995computer,torii2006fluid}. 
Nevertheless they are known to be time and memory consuming and
therefore most of them are restricted to local positions (i.e. single vessel, confluences) or a few vessel segments.
Although modeling techniques and computational efficiency are constantly improved,
a 3D simulation of the FSI in a large network of compliant vessels is still prohibitive.  
Reduced models have been derived by taking advantage of the physics of the blood flow in large vessels. 
If we assume an axisymmetric circular velocity profile in the vessel,
the 3D problem can be reduced to a 2D problem.
If we further assume that the wavelength is large compared to the radius of the vessel, a 1D model can be obtained.    
The 1D model is specially interesting for several reasons. First, this model captures well the behaviours of pulse wave, 
from which one can extract a lot of useful information about the
cardiovascular system. For example, the Pulse Wave Velocity (PWV) has been recognized by European
Society of Hypertension as a very important marker to the diagnosis and treatment of
hypertension~\cite{blacher1999aortic,mancia20072007}. Second, it
allows fast numerical computation, which permits real-time applications for medical planning.
Third, it also provides pertinent boundary conditions for 3D simulations
in multi-scale models~\cite{formaggia2001coupling,nobile2009coupling}.

The 1D model consists of a system of two partial differential equations (PDEs) for the conservation of mass
and momentum. The PDEs involve the flow rate $Q$, the
cross-sectional area $A$ and the average pressure $P$.
To close the system, the constitutive relation of the arterial wall which relates $P$ and $A$ is necessary.
After the insertion of this relation into the PDEs, a nonlinear hyperbolicity-dominated system is obtained.
Depending on the details of the modeling, there may be some additional terms.
Diffusive terms can appear due to an additional fluid viscous term~\cite{hughes1973one,vignon2004outflow}
or/and the wall viscoelasticity~\cite{formaggia2003one}.
The axial pre-stress of the wall or/and the wall inertia can lead to dispersive operators~\cite{formaggia2003one}. 

In case of weak nonlinearity (i.e. small perturbation around
the equilibrium state~\cite{lighthill2001waves,pedleyfluid}), we can linearize
the 1D governing equations and find analytical
solutions in frequency domain~\cite{nicoud2005numerical,wang2004wave}.
But for the full nonlinear system, analytical solutions are not available yet.
Thus several numerical schemes have been proposed and used to solve the system in time domain.
We roughly classify them in:
\begin{itemize}
\item Finite Difference (FD)
~\cite{elad1991numerical,olufsen2000numerical,pullan2002anatomically,reymond2009validation,saito2011one,Stergiopulos1992,zagzoule1986global}
\item Finite Volume (FV)
~\cite{cavallini2008finite,delestre2012well,Wibmer2004}
\item Finite Element (FE)
~\cite{alastruey2011pulse,formaggia2003one,malossi2012two,sherwin2003computational,vignon2004outflow}
\item Discontinuous Galerkin (DG)
~\cite{alastruey2011pulse,marchandise2009numerical, matthys2007pulse,mynard20081d,sherwin2003computational}
\end{itemize}

These schemes have been successfully applied in other communities
where researchers have to solve similar hyperbolic problems. For instance,
the MacCormack scheme (FD) was principally designed for gas dynamics (i.e. 1D
compressible Euler equations) and it was then successfully used to
compute blood flow~\cite{elad1991numerical,fullana2009branched}.
From ideas frequently applied in shallow water equations,
Delestre et al. obtained ``well balanced" schemes which properly treat the source term
induced by a tapered artery~\cite{delestre2012well}. 
The 1D model and the numerical solutions have been
validated by \textit{in vitro} experimental~\cite{alastruey2011pulse,saito2011one,wang2012comparing}
or \textit{in vivo} clinical data~\cite{devault2008blood,olufsen2000numerical,reymond2011validation,reymond2009validation,stettler1981theoretical,steele2003vivo}.
But usually only one particular scheme was chosen in a study and no cross comparisons among the schemes can be found.
Sherwin et al. presented a Taylor-Galerkin (FE) and a DG method in reference~\cite{sherwin2003computational}. 
The results of the two methods agree very well in a test case of an idealized vessel implanted with a stent.
But no further detailed comparisons were made. 
Moreover, their work considered an elastic arterial wall instead of a viscoelastic one.
In fact, the diffusive term induced by the viscoelasticity needs careful treatment.  
To our knowledge, there are no discussions in literature on the advantage/drawback 
of each scheme for a viscoelastic model.

Our objective in this paper is to make a cross comparison of the four numerical integration schemes and
to suggest the suitable conditions of application for each scheme.
In general, we note that FD schemes are not flexible enough
to treat complex computational geometries in high dimensions (2D or
3D). However, FD, FE and FV schemes of low order accuracy are in fact completely equivalent for 1D linear problems.  
But for problems with large nonlinearities, solutions with sharp gradient may appear and 
the performances of different schemes could be different. 
Equally important is the numerical accuracy. For DG scheme it may be tuned either by the degree
of the polynomial or by the mesh size. 
But if a diffusive term is added to the governing equations, 
the term will be hard to treat by an implicit time marching method (e.g. Crank-Nicolson) in the DG setting,
thus the time step may be very severely limited.  
Therefore, the performance of each scheme depends on the main features of the studied problems. 
In fact, the problems with different main features arise in a wide range of applications.
For instance, no shock is observed in arteries in normal physiological conditions 
but shock-like phenomena may arise in veins~\cite{flaud2012experiments,marchandise2010accurate,brook1999numerical} or 
in arteries when the human body suffers from a blunt impact by accident~\cite{kivity1974nonlinear}.
For another instance, in some conditions diffusive terms or dispersive terms may arise as source terms
~\cite{alastruey2011pulse} and the proper treatment of these terms will pose different levels of difficulty in each numerical framework.
Thus to make a cross comparison of the numerical schemes is interesting and useful.

 In this paper, Section~\ref{mathModel} presents the governing equations and
 the characteristic structure of the homogeneous part of the nonlinear
 system. Section~\ref{numSolver} describes the numerical solvers.
 In particular, a large amount of details of computation are given because this kind of information is
 scattered in literature. In this section, firstly an
 operator splitting is proposed (in the FD, FV and FE frameworks) to separate the hyperbolic and parabolic parts. 
 Then the treatment of the boundary conditions is discussed. Following that, MacCormack, Taylor-Galerkin 
 and MUSCL schemes are presented to integrate the hyperbolic subproblem.
 The parabolic subproblem is treated by a Crank-Nicolson method.
 At the end of this section, a local discontinuous Galerkin method is presented for the hyperbolic-parabolic problem without splitting. 
 Section~\ref{resultsDiscu} shows the analytical solutions and numerical results of the proposed schemes.
 The system is linearized and asymptotic solutions are obtained with different
 source terms in the system. 
 The effects of the skin friction and the viscosity of the wall on the pulse wave are clearly observed.
 Moreover, a wave with a step jump is computed and  
 the ability of the four schemes to properly capture the shock-like phenomena is tested.
 After that, a simple bifurcation is computed 
 and the numerical reflection and transmission coefficients are compared with the
 analytical ones predicted using linearized equations.
 Finally, a network with 55 arteries is computed.
 All the numerical solutions are compared favorably with the analytical,
 semi-analytical solutions or clinical observations.
In the last section, comparisons among the four schemes are made in four important aspects:
accuracy, ability to capture shock-like phenomena, computational speed
and implementation complexity. 
The suitable conditions for the application of each scheme are discussed.


\section{The 1D model of arterial blood flow}
\label{mathModel}
\subsection{1D mathematical model} 
The details of the derivation of the 1D model can be found in literature, such as~\cite{barnard1966theory,formaggia2009cardiovascular,hughes1973one,lagree2000inverse}.
We stress the two main assumptions usually held in most applications:
axisymmetric velocity profile and large wave length compared with the radius of the vessel.
The 1D arterial blood flow model can be written as:
\begin{subequations}
\begin{align} \frac{\partial A}{\partial t}+\frac{\partial Q}{\partial
x}=0,
\label{massConserv_AQ}\\ \frac{\partial Q}{\partial
t}+\frac{\partial}{\partial x}\left(\alpha
\frac{Q^2}{A}\right)+\frac{A}{\rho}\frac{\partial P}{\partial x}=
-C_f\frac{Q}{A} \label{momentumConserv_AQ},
\end{align} 
\end{subequations}
where as stated above, $A$ is the cross-sectional area of the
artery, $Q$ the volumetric flow rate or flux and $P$ the internal pressure.
The blood density $\rho$ is assumed a constant.
The independent variable $t$ is time and $x$ is the axial distance.
The coefficient $\alpha$ is the momentum correction factor, 
and  $C_f$ is the skin friction
coefficient. They depend on the shape of the velocity profile.
Usually, the profile can be estimated from the Womersley number which is defined as
$R\sqrt{\omega/ \nu}$, with $R$ the radius of the vessel, $\omega$ the angular frequency of the pulse wave and
$\nu$ the kinematic viscosity of the fluid. 
With a small Womersley number, we can take a Poiseuille (parabolic) profile.
In that case $\alpha=\frac{4}{3}$ 
and $C_f=8\pi\nu $. 
This choice is only valid for very viscous flows~\cite{lagree2000inverse, lagree1996etude}. 
In practice, viscosity is not so large, and the profile is more flat.
For a completely flat profile  $\alpha$
equals 1.
This value is often used since it leads to a
considerable simplification in analysis and the loss of relevance of
the model is very small in most cases~\cite{formaggia2003one}. Thus we
assume its value is 1 in this paper.  
The value of $C_f$ needs special attention because it has significant influence on the pulse wave.
In practical applications, its value has to be determined according to the particular problem at hand
(both \textit{in vitro} and \textit{in vivo} ones).
We assume its value is $8\pi\nu$ according to a Poiseuille profile.
We are aware of the limit of this approximation.  
However, as our  purpose is comparison of numerical schemes, we do not discuss any more the values of $\alpha$ and $C_f$.

To close the system, several  viscoelastic constitutive relations for
arterial wall have been presented in literature,
like~\cite{alastruey2011pulse,armentano1995arterial,holenstein1980viscoelastic,raghu2011comparative}.
We choose the Kelvin-Voigt model for simplicity\cite{alastruey2011pulse,armentano1995arterial}.
We assume that the arterial wall
is thin, isotropic, homogeneous, incompressible, and moreover that
it deforms axisymmetrically with each circular cross-section
independently of the others.
We denote the undeformed cross-sectional area by $A_0$ and the external pressure of the vessel by $P_{ext}$.
Then, the relation linking $A$ and $P$ is:
\begin{equation} P=P_{ext}+\beta(\sqrt{A}-\sqrt{A_0})+\nu_s \frac{\partial
A}{\partial t},\label{constitutive}
\end{equation} 
with the stiffness coefficient $\beta$,
\begin{equation*} \beta=\frac{\sqrt{\pi} Eh}{(1-\eta^2)A_0},
\end{equation*} and the viscosity coefficient $\nu_s $,
\begin{equation} \nu_s =\frac{\sqrt{\pi} \phi h}{2(1-\eta
^2)\sqrt{A_0}A},
\label{nu_Voigt}
\end{equation}
 where $\eta$ is the Poisson ratio, which is 0.5 for an
incompressible material, $E$ the Young's modulus, 
$h$ the thickness of the wall and $\phi$ the
viscosity of the material. For convenience,
we further define $C_v=\frac{A\nu_s }{\rho}$ for reasons
which will be clear very soon in the next section.
We also note that in absence of the wall viscosity we retrieve the
classical Hooke's law.

\subsection{Characteristic structure of the system}
After presenting the system of equations, we remind its hyperbolic feature by discussing the characteristic structure.
The discussion is classical, and can be found in text books~\cite{formaggia2009cardiovascular,leveque2002finite}.
The notations we introduce here will be useful for the discussion of the numerical solvers.
 We assume $P_{ext}$ is
constant along the axial variable $x$, and substitute the constitutive
relation~(\ref{constitutive}) into Eq.~(\ref{momentumConserv_AQ}).
We note that $\frac{\partial A}{\partial t}$ can be replaced
by $-\frac{\partial Q}{\partial x}$ thanks to Eq.~(\ref{massConserv_AQ}).
The equation for the balance of momentum turns out to
\begin{equation}
\frac{\partial Q}{\partial t}+
\frac{\partial}{\partial x}\bigl(\frac{Q^2}{A}+\frac{\beta}{3\rho}A^\frac{3}{2}\bigr)
-\frac{A}{\rho}\frac{\partial}{\partial x}\bigl(\nu_s\frac{\partial Q}{\partial x}\bigr)=
-C_f\frac{Q}{A}+ \frac{A}{\rho}\bigl(\frac{\partial (\beta\sqrt{A_0})}{\partial x} 
-\frac{2}{3}\sqrt{A}\frac{\partial \beta}{\partial x}\bigr).
 \label{momentum2}
\end{equation}
Under the assumption of a small perturbation of $A$, we approximate the term
$\frac{A}{\rho}\frac{\partial}{\partial x}(\nu_s\frac{\partial Q}{\partial x})$ by
$C_v\frac{\partial^2 Q}{\partial x^2}$
with the already defined coefficient
$C_v=\frac{A\nu_s }{\rho}=\frac{\sqrt{\pi} \phi h}{2\rho(1-\eta^2)\sqrt{A_0}}$,
which turns out to be independent of $A$ or $Q$.
The governing equations may be written as:
\begin{equation}
\frac{\partial U}{\partial t}+\frac{\partial F}{\partial x}=S, \label{vectorform1}
\end{equation}
where
\begin{equation*} 
U=\binom{A} {Q}, \quad
F=F_c+F_v=\binom{Q}{\frac{Q^2}{A}+\frac{\beta}{3\rho}A^\frac{3}{2}}+\binom{0}{-C_v\frac{\partial  Q}{\partial x}} 
\end{equation*}
and
\begin{equation*}
S=\binom{0}{-C_f\frac{Q}{A}+
\frac{A}{\rho}\bigl(\frac{\partial (\beta\sqrt{A_0})}{\partial x}
-\frac{2}{3}\sqrt{A}\frac{\partial \beta}{\partial x}\bigr)}.
\end{equation*}
In this equation, $U$ is the conservative variable, $F$ the corresponding flux
and $S$ the source term. 
Note that the flux (scaled by constant density) consists of two 
parts, the convective $F_c$ and the diffusive $F_v$. 
We recognize  $\frac{Q^2}{A}$ due to the fluid flow,  $\frac{\beta}{3\rho}A^\frac{3}{2}$ due to the elasticity, 
and $-C_v\frac{\partial Q}{\partial x}$ due to the viscosity of the wall.
In general, the suitable numerical techniques for the convective
and diffusive fluxes are different. Thus it is common to separate the
diffusive term and put it on the right side.
Thus we may write the problem in a convection-diffusion form:
\begin{equation}
\frac{\partial U}{\partial t}+\frac{\partial F}{\partial x}=S+D	\label{vectorform2}
\end{equation} 
with
\begin{equation*} 
F=F_c
,\quad
D=\binom{0}{
C_v\frac{\partial ^2 Q}{\partial x^2}}.
\end{equation*}
We consider firstly the homogeneous part and later the non-homogeneous part.
Expanding the derivative of the flux, the homogeneous part can be written in a quasi-linear form 
\begin{equation}
\frac{\partial U}{\partial t}+J_c\frac{\partial U}{\partial x}=0, \label{quasiform}
\end{equation}
where $J_c$ is the Jacobian matrix
\begin{equation*}
J_c=\begin{pmatrix}
0 & 1 \\
\frac{Q^2}{A^2}+c^2 &2\frac{Q}{A}
\end{pmatrix}
\end{equation*}
with the Moens-Korteweg celerity
\begin{equation}
c=\sqrt{\frac{\beta}{2\rho}A^{\frac{1}{2}}}.
\end{equation}
Actually, $A$ is always positive. Therefore $c$ is real,
which is the speed of the pressure wave with respect to the fluid flow.
The matrix $J_c$ has two different eigenvalues 
\begin{equation}
\lambda_{1,2}=\frac{Q}{A}\pm c.  
\end{equation}
Linear algebra shows $J_c$ must be diagonalizable in the 
form $J_c=R\Lambda R^{-1}$. The columns
of R are the right eigenvectors of $J_c$.
Left multiplying Eq.~(\ref{quasiform}) by $R^{-1}$, and introducing a new vector $W$ which satisfies $\partial_U W=R^{-1}$, one obtains
\begin{equation}
\frac{\partial W}{\partial t}+\Lambda \frac{\partial W}{\partial x}=0. \label{characterform}
\end{equation}
$W_{1,2}$ can be readily obtained by integrating $\partial _U W=R^{-1}$
componentwise
\begin{equation}
W_{1,2}=\frac{Q}{A}\pm 4c. \label{characteristics}
\end{equation}
$W=[W_1,W_2]^T$ is called Riemann invariant vector or characteristics.
In time-space plane, $W_{1,2}$ are constants along the lines $D_t X_{1,2}(t)= \lambda_{1,2}$.
In physiological conditions, $\lambda_1 >0> \lambda_2$.
The two families of characteristic propagate in opposite directions.
The homogeneous part is a subcritical hyperbolic system.
For further use, we get the expressions for A and Q 
by inverting the relation~(\ref{characteristics}), 
\begin{equation}
A=\frac{(W_1-W_2)^4}{1024} \left(\frac{\rho}{\beta}\right)^2, \quad
Q=A\frac{W_1+W_2}{2}.   \label{inverseCharac}
\end{equation}

In the non-homogeneous part, the skin friction term dissipates the momentum and
the second order derivative of $Q$ is diffusive.
Thus the full system has hyperbolic-parabolic features.
In physiological conditions, the Womersley number is
not too big and the artery is almost uniform, thus the source term will be very small
and the system is dominated by the hyperbolicity feature.
If the properties of the artery have sharp variations, large source terms will be introduced.
In this case, we will treat the artery as different segments connected together.

\section{Numerical solvers}
\label{numSolver}
Having defined the problem and notations, in this section we present the numerical solvers.
The original problem is split
into two subproblems which are respectively hyperbolic and parabolic. Three numerical schemes are presented to treat the hyperbolic 
subproblem.
For the parabolic  subproblem, 
Crank-Nicolson method is suitable.
Because of the duplication of values at the interfaces of elements in the DG setting,
there are difficulties to apply Crank-Nicolson scheme.
A local discontinuous Galerkin method is adopted to 
treat the problem without splitting.

\subsection{Operator splitting}
There are explicit high resolution schemes for hyperbolic problems. 
But for parabolic problems, implicit schemes are necessary in general
for a reasonable time step for time integration. Thus we applied a fractional step or operator 
splitting method. 
Starting from Eq.~(\ref{vectorform2}), the original problem is
split into to a hyperbolic subproblem, 
\begin{equation}
\frac{\partial U}{\partial t}+\frac{\partial F}{\partial x}=S   \label{subproblemHyper}
\end{equation} 
and a parabolic one,
\begin{equation}
\frac{\partial U}{\partial t}=D.     \label{subproblemPara}
\end{equation} 
Let us consider the time intervals $(t^n,t^{n+1})$, for $n=0,1,...,$ with 
$t^n=n\Delta t$.
In every time interval, the hyperbolic problem is solved to get a predictor $U^*$,
which is used as the initial condition (I.C.) of the second problem. The second 
step can be viewed as a corrector. The original problem is approximated by
a sequential application of the two subproblems in a certain order.

From data $U^n$,
we may make a prediction $U^*$ by evolving time $\Delta t$ of the hyperbolic subproblem,
and correct it with the evolution over $\Delta t$ of the parabolic subproblem,
\begin{equation*}
U^n \xrightarrow{e^{\Delta t\mathcal{H}}}U^*\xrightarrow{e^{\Delta t\mathcal{P}}}U^{n+1},
\end{equation*}
where $e^{\Delta t\mathcal{H}}$ ($e^{\Delta t\mathcal{P}}$)
means to solve the hyperbolic (parabolic) subproblem over $\Delta t$.
This method is called Godunov splitting. 
If the two subproblems are not commutable, the splitting error
is $\mathcal{O}(\Delta t)$, 
see Chapter 17 of reference~\cite{leveque2002finite}.
  
There is a 3-stage splitting called Strang splitting,
which has a leading error term $\mathcal{O}(\Delta t^2)$,
\begin{equation*}
U^n \xrightarrow{e^{\frac{1}{2}\Delta t\mathcal{P}}}U^*\xrightarrow{e^{\Delta t\mathcal{H}}}U^{**}
\xrightarrow{e^{\frac{1}{2}\Delta t\mathcal{P}}}U^{n+1}.
\end{equation*}
But in most cases the errors induced by the two splittings are very close.
That is because the coefficient of the term $\mathcal{O}(\Delta t)$ is much 
smaller then the coefficient of $\mathcal{O}(\Delta t^2)$~\cite{leveque2002finite}. 
We will see in Section~\ref{sec:resultDiffusion} a test case on the diffusion term.
The results show that the Godunov splitting is sufficient for our problem. 

Because the system is dominated by the hyperbolicity,
it must be driven mainly by the boundary conditions (B.C.) through the first subproblem. 
Thus we discuss the B.C. of the  
hyperbolic part in the next subsection and present the treatment of B.C. 
for the parabolic part in Section~\ref{numericalParabolic} together with
Crank-Nicolson scheme.

\subsection{Initial and  boundary conditions}
\subsubsection{Initial conditions}
Assume we are interested in the blood flow in an arterial segment $(0, L)$
within a time interval $(0, T)$. For an 
evolutionary problem, a proper I.C. is needed. 
In reality, the information contained in I.C. flows out after a 
certain interval of time, and it will not have influence on the system thereafter.
Thus, the I.C. can be set arbitrarily, say, 
$U(t=0,x)=(A_0,0)$, for convenience.

\subsubsection{Inlet and outlet of the homogeneous hyperbolic part} 

Assuming the source terms are small, we can impose the B.C. approximately by
taking advantage of the characteristic structure of the homogeneous part~\cite{formaggia2003one}.
Let us look back to the vector Eq.~(\ref{characterform}) again. 
The two components of this system are
\begin{subequations}
\begin{align} \frac{\partial W_1}{\partial t}+\lambda_1\frac{\partial
W_1}{\partial x}(U)=0,
 \label{seperateCharacForm1} \\
\frac{\partial W_2}{\partial t}+\lambda_2\frac{\partial W_2}{\partial
x}(U)=0. \label{seperateCharacForm2}
\end{align}
\end{subequations} 
Since the two eigenvalues have opposite signs, there is
exactly one incoming characteristic at each end of the computational
domain. The incoming characteristic carries information from outside of
the domain and thus is essential to guarantee the problem to be
well-posed. That is to say, the system must be supplemented by B.C.s in the form
\begin{equation} 
W_1(0,t)=g_1(t), \quad W_2(L,t)=g_2(t), t>0.
\label{incomingCharacteristics}
\end{equation}

The outgoing characteristic carries information from inside of the
domain, which can be given by the differential equations. 
Since $W_{1,2}$ are constants along the lines
 $D_t X_{1,2}(t)= \lambda_{1,2}$ in time-space plane, we can get $W_2^{n+1}(0)$
 and $W_1^{n+1}(L)$ by interpolation in the data of the $n$-th time step:
\begin{equation} W_2^{n+1}(0)=W_2^n\bigl(-\lambda_2 ^n(0)\Delta
t\bigr), \quad W_1^{n+1}(L)=W_1^n\bigl(L-\lambda_1 ^n(L)\Delta
t\bigr). \label{extroplation}
\end{equation} 
The characteristics are then transformed to physical
variables by relation~(\ref{inverseCharac}) for numerical computation.

In reality, we rarely have the explicit expression~(\ref{incomingCharacteristics}) for the incoming characteristics.
Usually, we want to impose B.C. in physical term $A$, $Q$ or $P$.
At the inlet, if $A^{n+1}$ is given, one can use the
relation~(\ref{characteristics}) to deduce:
\begin{equation*}
W_1^{n+1}=W_2^{n+1}+8\sqrt{\frac{\beta}{2\rho}\sqrt{A^{n+1}} }.
\end{equation*} 
If $Q^{n+1}$ is given, we approximate $A^{n+1}$ by $A^n$ and then obtain
\begin{equation*}
W_1^{n+1}=-W_2^{n+1}+2\frac{Q^{n+1}}{A^{n}}.
\end{equation*}
If $P^{n+1}$ is given, from the wall relation (\ref{constitutive}) simplified with no viscous effect ($\nu_s=0$),
 we in fact impose:
\begin{equation*}
W_1^{n+1}=W_2^{n+1}+8\sqrt{\frac{1}{2\rho}(P^{n+1}+\beta A_0^{1/2})}.
\end{equation*} 
At the outlet, some part of the perturbation of outgoing characteristic
$W_1$ may be reflected,
\begin{equation*} 
W_2^{n+1}=W_2^{0}-R_t(W_1^{n+1}-W_1^{0}),
\end{equation*} 
where $R_t$ is the coefficient of reflection. If $R_t=0$, the B.C. is 
nonreflecting. That means the outgoing characteristic goes out without
leaving any effect and that the incoming characteristic is a constant in time.
If there are changes of properties in the downstream of the vessel, usually a nonzero 
$R_t$ will be incurred. 

\subsubsection{Conjunction points}

There are many cases when conjunctions of different vessels need to be considered: when there are changes of topology,
sharp variations in geometrical or mechanical properties. 
Topological changes correspond to the large amount of bifurcations and some trifurcations in the arterial network. 
Sharp variations may also arise in many conditions, for example when there are increases of stiffness $\beta$ due to stenting or $A_0$ due to aneurysm.
In these cases, the derivatives of the corresponding variables in the source terms are very large or even near a singularity,
and then the vessel can be treated as several joined segments with different properties. 

Since all of the conjunction points can be treated with the same method, we consider
a branching point as a sample problem: a parent vessel with two daughter
arteries. At the branching point, there are then six boundary
conditions, $A_p^{n+1}$ and $Q_p^{n+1}$ for the outlet of the parent
artery and $A_{d_1}^{n+1}$, $Q_{d_1}^{n+1}$,$A_{d_2}^{n+1}$ and
$Q_{d_2}^{n+1}$ for the inlets of the two daughter arteries.  From the physical
point  of view, we have to preserve the conservation of
mass flux
\begin{subequations}
\begin{equation}
Q_p^{n+1}-Q_{d_1}^{n+1}-Q_{d_2}^{n+1}=0,     \label{equ1}
\end{equation}
and conservation of momentum flux
\begin{equation}
\frac{1}{2}\rho\left(\frac{Q_p^{n+1}}{A_p^{n+1}}\right)^2+P_p^{n+1}-\frac{1}{2}\rho\left(\frac{Q_{d_i}^{n+1}}{A_{d_i}^{n+1}}\right)^2-P_{d_i}^{n+1}=0
\quad i=1,2.  \label{equ2}
\end{equation}
The pressures $P_p^{n+1}$ and $P_{d_i}^{n+1}$ shall be expressed in cross-sectional area $A$ by the constitutive relation~(\ref{constitutive}).
In the Eqs.~(\ref{equ2}), there may be some terms for energy losses due to the branching~\cite{steele2003vivo,formaggia2003one,matthys2007pulse}.
But in practice, these losses only have secondary effects on the pulse waves~\cite{matthys2007pulse}. 
Therefore we did not include them.  
 
Moreover, the outgoing characteristics of the joined arteries should be matched.
In the parent artery, $(W_1)_p^{n+1}$ is given by the data on the $n$-th 
time step with the interpolation formula~(\ref{extroplation}).
It must be equal to $W_1(U_p^{n+1})$
which is given by relation~(\ref{characteristics}).
Thus we have the equation  
\begin{equation}
(W_1)_p^{n+1}-W_1(U_p^{n+1})=0. \label{equ3} 
\end{equation} 
The same principle holds for $W_2$ on the two daughter arteries,
\begin{equation}
(W_2)_{d_i}^{n+1}-W_2(U_{d_i}^{n+1})=0 \quad i=1,2.   \label{equ4}
\end{equation}
\end{subequations}
Combining Eqs.~(\ref{equ1}),~(\ref{equ2}),~(\ref{equ3}) and~(\ref{equ4}),
there are 6 Eqs. with 6 unknowns. 
This nonlinear algebraic system can be readily solved by Newton-Raphson iterative method with $U^n$ as the initial guess.
In our test, the computation converges very fast.
Usually a very few iterations are enough for a satisfactory accuracy.

\subsection{MacCormack scheme}

In FD framework, MacCormack method~\cite{Maccormack69}
is very suitable for nonlinear hyperbolic systems of conservation laws.
It is equivalent to the Lax-Wendroff scheme for linear systems.
It has the following characteristics: conservative form, 
three-point spatial stencil and two time levels (predictor and corrector), 
second-order accuracy in time and space.

\begin{figure}[ht] \center
\includegraphics[width=0.9\textwidth]{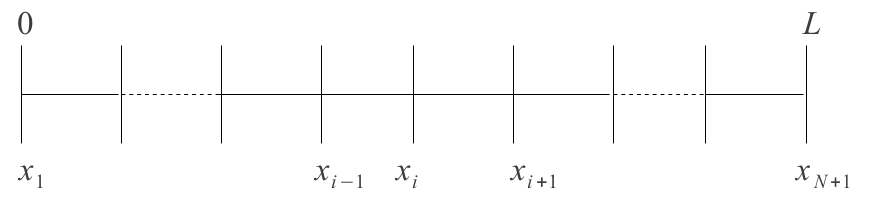}
\caption{Mesh for FD and FE}
\label{meshFD}
\end{figure}

The numerical solution is
performed in a mesh with $N+1$ points and thus the spatial resolution is $\Delta x = { L \over N}$, see Figure~\ref{meshFD}. 
For the conservative system~(\ref{subproblemHyper}), an 
approximate solution $\bold{U}^{*}$ is obtained first from $\bold{U}^n$ 
and then $\bold{U}^{*}$ is corrected to give the solution  $\bold{U}^{n+1}$ at
the time step $t+\Delta t$. The finite difference equations  (at
the interior grid points) are then :
\begin{enumerate}
\item predictor step
  \begin{equation*} U_i^{*}=U_i^{n}-\frac{\Delta t}{\Delta
x}(F^n_{i+1}-F^n_{i})+\Delta t S^n_{i},
\quad 
i = 2,...N
  \end{equation*}
\item corrector step
  \begin{equation*}
U_i^{n+1}=\frac{1}{2}(U_i^{n}+U_i^{*})-\frac{\Delta t}{2\Delta
x}(F^*_{i}-F^*_{i-1})+\frac{\Delta t}{2} S^*_{i},\quad 
i = 2,...N
  \end{equation*}
\end{enumerate}
where $\bold{F}^*$ and $\bold{S}^*$ are evaluated as functions of the predicted solution $\bold{U}^*$.
Note that the predictor step applies a forward differencing and the corrector step a backward differencing.
The order of the two kinds of differencing can be reversed. 
The grid points $x_1$ and $x_{N+1}$ represent the boundary conditions.

\subsection{Taylor-Galerkin scheme} 
In this section, we follow the presentations of Formaggia et al.~\cite{formaggia2003one,formaggia2009cardiovascular}
and Sherwin et al.~\cite{sherwin2003computational} for the Taylor-Galerkin 
scheme. Other forms are also possible, see the reference~\cite{wan2002one} for example. 

From Eq.~(\ref{subproblemHyper}), one may obtain,
\begin{equation} \frac{\partial U^n}{\partial t}=S^n-\frac{\partial
F^n}{\partial x} .  \label{1stOrder}
\end{equation}
Differentiating both sides with respect to $t$ and exchanging the order of spatial 
and temporal differentiations in the second term give
\begin{equation} \frac{\partial ^2 U^{n}}{\partial t^2}=\Big(S_U
\frac{\partial U}{\partial t}\Big)^n -\frac{\partial}{\partial x}\Big(
H\frac{\partial U}{\partial t}\Big)^n,   \label{2stOrder}
\end{equation} 
where $S_U=\frac{\partial S}{\partial U}$ and $H=\frac{\partial F}{\partial U}$.  
Substituting Eq.~(\ref{1stOrder}) into Eq.~(\ref{2stOrder}) and then both of 
them into the Taylor series of $U^{n+1}$ up to the second order, one gets, 
\begin{equation}
\begin{split} U^{n+1}=U^n-\Delta t\frac{\partial}{\partial x}\bigl[
F^n+\frac{\Delta t}{2}H^nS^n \bigl] -\frac{\Delta
t^2}{2}\bigl[S_U^n\frac{\partial F^n}{\partial x}
-\frac{\partial}{\partial x}\bigl(H^n\frac{\partial F^n}{\partial
x}\bigr)\bigr] \\ +\Delta t\bigl(S^n+\frac{\Delta t}{2}S_U^nS^n\bigr).
\label{Taylorexpan}
\end{split}
\end{equation}
For convenience, we adopt the notations
\begin{align*} F_{LW}(U)=F(U)+\frac{\Delta t}{2}H(U)S(U), \\
S_{LW}(U)=S(U)+\frac{\Delta t}{2}S_U(U)S(U).
\end{align*} 
The piecewise linear function space associated with the mesh
(Figure~\ref{meshFD}) is given as
\begin{equation*} V_h^0=\lbrace [v_h]^2 |v_h \in \mathcal{C}^0,
v_h|_{[x_i,x_{i+1}]} \in \mathcal{C}^1, v_h(0)=v_h(L)=0, i=1...N\rbrace.
\end{equation*}
This is both the trial function 
space and the test function space in Galerkin framework.
We further define the 
inner product
\[
(U,V)=\int_0^L U \cdot V dx.
\]
At the interior points $x_2,...x_N$, if we approximate $U$ by $U_h \in V_h^0$ in Eq.~(\ref{Taylorexpan}), multiply both sides by 
basis test functions $\psi_i \in V_h^0$, and integrate over the domain $[0,L]$, finally we can get
\begin{equation}
\begin{split} (U_h^{n+1},\psi_i)=(U_h^n,\psi_i)+\Delta
t\bigl(F_{LW}(U_h^n),\frac{d\psi_i}{dx}\bigl) -\frac{\Delta
t^2}{2}\bigl(S_U(U_h^n)\frac{\partial F(U_h^n)}{\partial x},\psi_i\bigl)\\
-\frac{\Delta t^2}{2}\bigl(H(U_h^n)\frac{\partial F(U_h^n)}{\partial
x},\frac{d\psi_i}{dx}\bigl)+\Delta t(S_{LW}(U_h^n),\psi_i)  \label{weakform}
\end{split}
\end{equation} 

In computation, we enforce the Eq.~(\ref{weakform}) componentwise. That is,
\begin{align*}
(A_h^{n+1},v_i)=RHS1^n_i, \\
(Q_h^{n+1},v_i)=RHS2^n_i,
\end{align*}
where $ v_i $ is one component of the vector $ \psi_i $ and 
\begin{equation}
\begin{split}
RHS1^n_i=(A_h^n,v_i)+\Delta
t\Bigl( \bigl[F_{LW}(U_h^n)\bigr]_1,\frac{d v_i}{dx}\Bigr) -\frac{\Delta
t^2}{2}\Bigl( \bigl[ S_U(U_h^n)\frac{\partial F(U_h^n)}{\partial x} \bigr]_1  ,v_i\Bigr)\\
-\frac{\Delta t^2}{2}\Bigl( \bigl[H(U_h^n)\frac{\partial F(U_h^n)}{\partial
x}\bigr]_1,\frac{d v_i}{dx}\Bigr)+\Delta t\Bigl( \bigl[ S_{LW}(U_h^n)\bigr]_1, v_i \Bigr).  \label{weakformSeperat1} 
\end{split}
\end{equation} 
The form $[\cdot ]_1$ indicates the first component of the vector in the bracket.
$RHS2_i^n$ can be expressed in a similar way.

To elaborate the computing details, we take the Eq.~(\ref{weakformSeperat1}) as an example.
In the FE framework, $A_h^{n+1}$ and $A_h^{n}$ are expanded as $A_h=\sum_{j=2}^{j=N}A_j v_j$.
We denote the unknown vector $(A_2,\dots A_{N})^T$ by $\bold{A}$.
Instead of evaluated directly as nonlinear functions of $U_h^n$,
the terms $F(U_h^n)$, $F_{LW}(U_h^n)$, $S_{LW}(U_h^n)$, $S_U(U_h^n)$ and $H(U_h^n)$
are projected onto the trial function space and expanded by a group finite element method. 
That is, for example, $[F(U_h^n)]_1=\sum_{j=2}^{j=N} [F^n_j ]_1 v_j$ with $[F^n_j]_1=[F(U^n_j)]_1$.
Finally, the matrix form of Eq.~(\ref{weakformSeperat1}) writes
\begin{equation}
\begin{split}
 \mathcal{M}\bold{A}^{n+1}=\mathcal{M}\bold{A}^n+\Delta t \mathcal{K}^T [\bold{F}_{LW}^n]_1
-\frac{\Delta t^2}{2} (\tilde{\mathcal{M}_1} [\bold{F}^n]_1 +\tilde{\mathcal{M}_2} [\bold{F}^n]_2)  \\
  -\frac{\Delta t^2}{2} ( \tilde{\mathcal{K}_1} [\bold{F}^n]_1+\tilde{\mathcal{K}_2} [\bold{F}^n]_2)+\Delta t \mathcal{M} [\bold{S}_{LW}^n]_1 ,
\end{split}
\end{equation} where
\begin{equation*} \mathcal{M}_{ij}=( v_i, v_j), \quad
\mathcal{K}_{ij}= (v_i, \frac{\partial{v_j}}{\partial x} )
\end{equation*}
and
\begin{align*}
\tilde{\mathcal{M}_1}(S_u)_{ij}=\bigg(\sum_k (S^{(1,1)}_u)_k v_k \frac{\partial v_i}{\partial x}, v_j\bigg), \quad 
\tilde{\mathcal{M}_2}(S_u)_{ij}=\bigg(\sum_k (S^{(1,2)}_u)_k v_k \frac{\partial v_i}{\partial x}, v_j\bigg), \\
\tilde{\mathcal{K}_1}(H)_{ij}=\bigg(\sum_k H^{(1,1)}_k v_k \frac{\partial v_i}{\partial x}, \frac{ \partial v_j}{\partial x}\bigg), \quad
\tilde{\mathcal{K}_2}(H)_{ij}=\bigg(\sum_k H^{(1,2)}_k v_k \frac{\partial v_i}{\partial x}, \frac{ \partial v_j}{\partial x}\bigg).
\end{align*}
The form $(S_u^{(\cdot, \cdot)})_k$ indicates the $k$-th component of the vector at the position $(\cdot,\cdot)$ of the discretized matrix $\bold{S}_u$. 
Please note that the operators $\tilde{\mathcal{M}_1}$ etc. are functions of $\bold{S}_u$ and $\bold{H}$, 
therefore they must be updated in every time step.

\subsection{MUSCL} 
\label{Finite_Volume}

\begin{figure}[ht] \center
\includegraphics[width=0.9\textwidth]{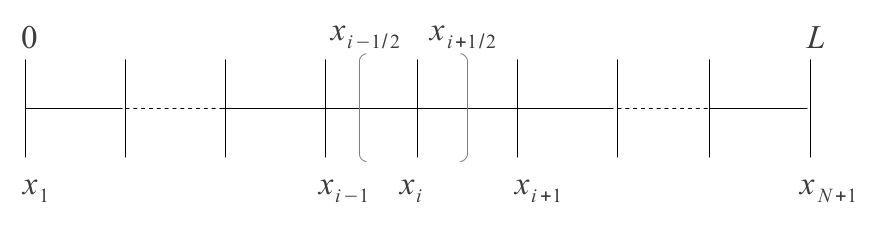}
\caption{Mesh for FV}
\label{meshFV}
\end{figure}

In this section, we mainly follow the presentation~\cite{delestre2012well} but with a different temporal integration method.
For finite volume method, the domain is decomposed into finite volumes or
cells with vertex $x_i$ as the center of cell $[x_{i-1/2}, x_{i+1/2} ]$, see Figure~\ref{meshFV}.
In each cell, average values are considered,
\begin{equation*} U_i=\frac{1}{\Delta
x}\int_{x_{i-1/2}}^{x_{i+1/2}}U(x)dx
,\quad
S_i=\frac{1}{\Delta
x}\int_{x_{i-1/2}}^{x_{i+1/2}}S(x)dx.
\end{equation*}
Integrating the governing equations over each cell and applying Gauss's theorem, one readily obtains 
\begin{equation}
\frac{dU_i}{dt}=-\frac{(F|_{x_{i+1/2}}-F|_{x_{i-1/2}})}{\Delta x}+S_i  .    \label{ODEofFV}
\end{equation}
We have a local Riemann problem at each interface of neighboring cells,
since $U_{i+1/2-}$ and $U_{i+1/2+}$, the left limit of $U_i$ and the right limit of $U_{i+1}$ at $x_{i+1/2}$ respectively, are not equal in general.
By solving the Riemann problem, a numerical flux $F^*$ can be obtained. 
Depending on the approximate approaches on solving the Riemann problem, 
different numerical fluxes are possible. 
Among them, Rusanov (or called local Lax-Friedrichs) flux is widely used.
According to reference~\cite{bouchut2004nonlinear}, it writes
\begin{equation*}
F_{i+1/2}^*=\frac{F(U_{i+1/2-})+F(U_{i+1/2+})}{2}-c\frac{U_{i+1/2+}-U_{i+1/2-}}{2},
\end{equation*} with
\begin{equation*} c=\max \big (\lambda_1(U_{i+1/2-}),\lambda_1(U_{i+1/2+}) \big),
\end{equation*} where $\lambda_1$ is the biggest eigenvalue of $J_c$. 
Other numerical fluxes with less numerical diffusivity are possible, such as HLL (Harten-Lax-Van Leer) flux~\cite{bouchut2004nonlinear,delestre2012well}.
Since Rusanov flux is more simple and robust, it is adopted 
in this paper.
If $\mathbf{U}_-$ and $\mathbf{U}_+$ are equal to the average values
at the cells, the scheme will be of first order accuracy. 
Reconstructions of $\mathbf{U_-}$ and $\mathbf{U_+}$ from $\mathbf{U}$ are necessary for a scheme of higher resolution. 

Let us consider the techniques of reconstruction.
For a scalar $s$ within the $i$-th cell, we denote its slope as $Ds_i$, which can be approximated
by $(s_i-s_{i-1})/\Delta x$, $(s_{i+1}-s_{i})/\Delta x$ or
$(s_{i+1}-s_{i-1})/2\Delta x$.
Then the values of $s$ at the interfaces associated with this cell can be recovered as
\begin{equation*} s_{i-1/2+}=s_i-\frac{\Delta x}{2} Ds_i, \quad
s_{i+1/2-}=s_i+\frac{\Delta x}{2} Ds_i.
\end{equation*} 
The discretization of derivative in space can achieve a second order accuracy by 
this method. 
But the solution will have  nonphysical oscillations. Some
examples of oscillations induced by these methods can be found in 
Chapter 6 of reference~\cite{leveque2002finite}. Slope or flux limiter 
and non-oscillatory solutions are integral characteristics of FV schemes.
MUSCL (monotonic upwind scheme for conservation law) is one popular slope limited 
linear reconstruction technique.
To present MUSCL, we first define a slope limiter, \begin{equation*} \text{minmod(x,y)}=
\begin{cases} \text{min(x,y)}& \text{if $x,y \geq 0$},\\
\text{max(x,y)}& \text{if $x,y \leq 0$},\\ 0& \text{else}
\end{cases}
\end{equation*}
Then the slope $Ds_i$ is modified as
\begin{equation*}
Ds_i=\text{minmod}(\frac{s_i-s_{i-1}}{\Delta x},\frac{s_{i+1}-s_{i}}{\Delta x}).
\end{equation*}
The values of $\mathbf{U_-}$ and $\mathbf{U_+}$ at the interfaces can be obtained by linear reconstruction with the slope $Ds_i$.
The variables are conserved by this reconstruction.

After the discretization in space, we have the semi-discrete form,
\begin{equation*}
\frac{dU_i}{dt}=\Phi(U_{i-2},...U_{i+2})
\end{equation*} 
where
\begin{equation*} 
\Phi(U_{i-2},...U_{i+2})=-\frac{(F^*_{i+1/2}-F^*_{i-1/2})}{\Delta x}+S_i.
\end{equation*} 
The numerical fluxes $F^*_{i+1/2}$ and $F^*_{i-1/2}$ are given by Rusanov flux 
with the reconstructed values at the two sides of the interfaces.
Note that this is a scheme with five stencils. 
The values at $x_1$ and $x_{N+1}$ are determined by the aforementioned characteristic method.
One ghost cell at each end of the computational domain is needed
and we approximate the values at these cells by the ones at the neighboring boundary cells.  

For the temporal integration, we may apply a 2-step second order Adams-Bashforth (A-B) scheme,  
\begin{equation*} 
\mathbf{U}^{n+1}=\mathbf{U}^n+\Delta t \left(\frac{3}{2}\Phi(\mathbf{U}^{n})-\frac{1}{2}\Phi(\mathbf{U}^{n-1})\right).
\end{equation*}
This scheme can be initiated by a forward Euler method.
Also, a second order Runge-Kutta (R-K) approach, namely Heun method is possible~\cite{shu1988efficient}.
It writes
\begin{align*}
\bold{U}^*=\bold{U}^n+\Delta t\Phi(\bold{U}^n), \\
\bold{U}^{**}=\bold{U}^*+\Delta t\Phi(\bold{U}^*), \\
\bold{U}^{n+1}=(\bold{U}^*+\bold{U}^{**})/2.
\end{align*}
Comparing the two methods, we note that $\Phi(\bold{U})$ has to be computed twice in R-K in every time step 
while the A-B method only needs once since $\Phi(\bold{U}^{n-1})$ is stored in the previous step and reused in the current step.   
Because the boundary conditions are determined dynamically to compute $\Phi(\bold{U})$, the R-K also incurs 
one more resolution of the nonlinear algebraic equations at conjunction points.   
For these reasons, we choose the A-B method for the temporal integration, although the R-K method usually allows
a larger time step size for convergence.

\subsection{Treatment of the parabolic subproblem}
\label{numericalParabolic}
For the previous 3 schemes, only the hyperbolic subproblem resulted from splitting is solved. 
For the parabolic subproblem, Crank-Nicolson method is very suitable.
The temporal and spatial discretization has the form,
\begin{equation*} \frac{U_i^{n+1}-U_i^*}{\Delta
t}=\frac{C_v}{2}\bigl(\frac{U_{i+1}^{n+1}-2U_i^{n+1}+U_{i-1}^{n+1}}{\Delta
x^2} +\frac{U_{i+1}^{*}-2U_i^{*}+U_{i-1}^{*}}{\Delta
x^2}\bigl),
\end{equation*} 
where $\mathbf{U}^*$ is the solution of the first hyperbolic subproblem.
The matrix of the resulting algebraic system is tridiagonal, which is quite
cheap to invert. This scheme is second order accurate both on time and space.  
Moreover, it is unconditionally stable.
It is natural to set a homogeneous Neumann B.C. for the parabolic subproblem,
$\partial _x U_p(0,t)=\partial _x U_p(L,t)=0$.
The subscript $p$ stands for parabolic.
We note that a second order implicit FE method can also be applied here.
But since this subproblem is linear and in 1D, the FE method would be exactly equivalent with this FD method.  

\subsection{Local Discontinuous Galerkin scheme}

\begin{figure}[ht] \center
\includegraphics[width=0.9\textwidth]{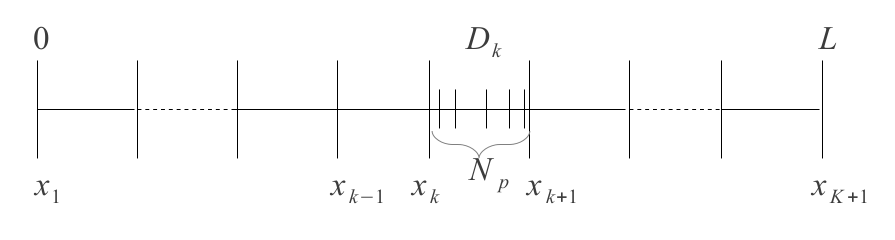}
\caption{Mesh for DG}
\label{meshDG}
\end{figure}

In the FV framework, the recovery of $\mathbf{U}_-$ and $\mathbf{U}_+$ of higher accuracy requires a big stencil.  In higher dimensions,
this kind of reconstruction leads to difficulties if the mesh is unstructured. On the other hand, it is quite straightforward to increase the 
order of approximation polynomials in one finite element. 
Unlike the global FE, the neighboring elements do not share the same values at
the interfaces. Numerical fluxes are obtained from these values, 
where the dynamics of the system can be considered. 
We present a nodal DG scheme, following Hesthaven and Warburton's book~\cite{hesthaven2008nodal}.     
The domain is decomposed into $K$ non-overlapping elements, see Figure~\ref{meshDG}. At each element, the local approximation to
the solution is a polynomial of order $N=N_p-1$. 
The global approximation to $U$ is the direct summation of these local solutions:
\begin{equation}
U_h=\bigoplus_{k=1}^{k=K}U_h^k.
\end{equation}
Similarly, the flux $F$ and the source term $S$ can also be  approximated
by the direct summation of piecewise $N$-th degree polynomials.
The local form of the conservation law on the $k$-th element is
\begin{equation}
 \frac{\partial U_h^k}{\partial t} +\frac{\partial F_h^k}{\partial x} =  S_h^k. \label{localIntegralForm}
\end{equation}
Multiplying both sides of Eq.~(\ref{localIntegralForm}) with a test function $\psi^k$, and
integrating over one element give
\begin{equation}
\bigg(\frac{\partial U_h^k}{\partial t}, \psi^k \bigg)_{D_k}+\bigg( \frac{\partial
F_h^k}{\partial x}, \psi^k \bigg)_{D_k}=\bigg( S_h^k, \psi^k \bigg)_{D_k}.
\end{equation} 
Applying integration by part on the second term, we have:
\begin{equation}
\bigg( \frac{\partial U_h^k}{\partial t}, \psi^k \bigg)_{D_k}- \bigg( F_h^k , \frac{\partial
 \psi^k}{\partial x} \bigg)_{D_k}+F_h^k\psi^k \bigg| _{x_{k}}^{x_{k+1}}=\bigg(S_h^k ,\psi^k \bigg)_{D_k}.
\end{equation} 
At the interface of $x_k$, the values of $U_h$ at the two sides, $U_h^{k-1}(x_k)$ and $U_h^k(x_k)$, are not guaranteed equal.
A numerical flux $F_k^*$ is introduced here. 
Through the numerical flux, information is communicated between elements.
In practice, the second term is integrated by part again for convenience 
of computation. Thus we have
\begin{equation}
\bigg( \frac{\partial U_h^k}{\partial t}, \psi^k \bigg)_{D_k}+\bigg( \frac{\partial
 F_h^k}{\partial x}, \psi^k \bigg)_{D_k} +\psi^k(-F_h^k + F^*) \bigg| _{x_{k}}^{x_{k+1}}
 =\bigg( S_h^k, \psi^k \bigg)_{D_k} \label{strongformDG}.
\end{equation} 
If we introduce $N_p$ nodes within the element $D_k$ (Figure~\ref{meshDG}),
the local solution can be expanded as
\begin{equation}
U_h^k(x,t)=\sum_{i=1}^{N_p}U_h^k(x_i^k,t)\ell_i^k(x),
\end{equation}
where $\ell_i^k(x)$ is the Lagrange interpolant  
associated with the $i$-th node. 
For the Galerkin scheme, Eq.~(\ref{strongformDG}) must hold for every test
function $\ell_i^k(x)$. 
Thus we have $N_p$ equations for $N_p$ unknowns.
In matrix form, the system can be written as,
\begin{equation}
\mathcal{M}^k\frac{d\bold{U}^k}{dt}+\mathcal{K}^k\bold{F}^k+\boldsymbol{\ell}^k (-F_h^k + F^*) \bigg| _{x_{k}}^{x_{k+1}}=\mathcal{M}^k\bold{S}^k,
\end{equation}
where
\[
\mathcal{M}^k_{(i,j)}=\big( \ell^k_i, \ell^k_j \big)_{D_k},
\quad
\mathcal{K}^k_{(i,j)}=\big( \ell^k_i, \frac{d\ell^k_j}{dx} \big)_{D_k},
\]
and $\boldsymbol{\ell}^k$ is the vector of functions $(\ell_1^k, \ell_2^k,..\ell_{N_p}^k)^T$.
The system of equations can be turned into a semi-discrete form,
\begin{equation}
\frac{d\bold{U}^k}{dt}=-\mathcal{D}^k\bold{F}^k + (\mathcal{M}^k)^{-1}\boldsymbol{\ell}^k (F_h^k - F^*) \bigg| _{x_{k}}^{x_{k+1}}
+\bold{S}^k ,
\end{equation}
where 
\[
\mathcal{D}^k_{(i,j)}=\Big( (\mathcal{M}^k)^{-1}\mathcal{K}^k \Big )_{(i,j)}=\frac{d\ell^k_j}{dr} \bigg|_{r_i}
\]
is the local differentiation operator~\cite{hesthaven2008nodal}.
The computation of $\mathcal{M}^k$ and $\mathcal{D}^k$ is crucial.
We define an affine mapping from
a reference element $(-1,1)$ to $D_k$,
\[
x(r)=x_k+\frac{1+r}{2}(x_{k+1}-x_k).
\]
The local operators can be readily computed as 
\[
\mathcal{M}^k_{(i,j)}=\mathcal{J}_k\int_{-1}^{1}\ell_i\ell_jdr,
\quad
\mathcal{D}^k_{(i,j)}=\mathcal{J}^{-1}_k\frac{d\ell_j}{dr} \bigg|_{r_i},
\]
where $\mathcal{J}_k=(x_{k+1}-x_{k})/2$ and $\ell_i$, $\ell_j$ are the Lagrange interpolants at the reference element. 
Note that the operators $\mathcal{M}^k$ and $\mathcal{D}^k$ can be precomputed
and stored. Legendre-Gauss-Lobatto points have to be chosen as the 
interpolation points to minimize computation error.
For more details, we refer to Chapter 3 of reference~\cite{hesthaven2008nodal}.
For the temporal integration, a second order A-B scheme is applied for reasons as discussed
in Section~\ref{Finite_Volume}.

The scheme previously presented can treat a hyperbolic problem.
But in this setting Crank-Nicolson method is hard to apply,
because the values at the interfaces are duplicated.
We consider the problem formulation 
of Eq.~(\ref{vectorform1}), where the flux contains convective part
$F_c$ and diffusive part $F_v$. For the convective part, Rusanov flux as mentioned in Section~\ref{Finite_Volume} is applicable. 
For the diffusive flux, a straight idea is to use the central flux, $(F_v(U_-)+F_v(U_+))/2$.
But as pointed out by Shu el al.~\cite{shu2001different}, this choice is inconsistent.

To solve this problem, we rewrite the original equations as
\begin{align*}
\frac{\partial U}{\partial t}+ \frac{\partial (F_c-C_vq)}{\partial x}=S \\
q-\frac{\partial Q}{\partial x}=0
\end{align*}
In semi-discrete form, the equations for one element are 
\begin{align*}
\frac{d\bold{U}_k}{dt}=-\mathcal{D}^k\bold{F}^k + (\mathcal{M}^k)^{-1}\boldsymbol{\ell}^k (F_h^k - F^*) \bigg| _{x_{k}}^{x_{k+1}}
+\bold{S}^k \\
\bold{q}^k=\mathcal{D}^k\bold{Q}^k-(\mathcal{M}^k)^{-1} \boldsymbol{\ell}^k (Q_h^k - Q^*) \bigg| _{x_{k}}^{x_{k+1}}
\end{align*}
The fluxes in these equations have to be modified accordingly: $\mathbf{F}^k=\mathbf{F}_c^k-C_v\mathbf{q}^k$,
$F_h^k=(F_c)_h^k-C_v q_h^k$ and $F^*=F^*_c-(C_vq)^*$. The convective flux $F^*_c$ is defined by Rusanov flux.
The fluxes $(C_vq)^*$ and $Q^*$ are defined by the central flux.
The introduction of an auxiliary variable $q$ stabilizes 
the scheme. Note that the auxiliary equation does not involve time evolution.
The computation and storage of $\bold{q}^k$ incur very limited extra costs.
This method is called local discontinuous Galerkin scheme.  

\section{Results and discussion}
\label{resultsDiscu}
The implementation codes can be verified by analytical solutions of linearized model or manufactured solutions of the full system without linearization~\cite{raghu2011verification,raghu2011comparative}.
In this paper, except comparisons with the homogeneous linearized model and results in literature, 
we derived asymptotic solutions with different source terms.
The verification by asymptotic analysis is a different approach from previous works. 
In this section, the computations are done on a single uniform vessel at first.
In case of small perturbations, a linearized system is obtained.  If this system is homogeneous, it allows pure wave solution.
If the source terms due to the skin friction and the viscosity of the wall are added respectively, asymptotic solutions are obtained.
In case of larger perturbations,  the full nonlinear system allows shocks.
The shock-capturing property of each scheme is tested in this case.  
After the tests on a single vessel, a simple bifurcation is computed and the reflection and transmission coefficients are compared
with analytical ones predicted by linearized system.
At the end of this section, a network with 55 arteries is computed and the numerical solutions are checked against clinical observations 
reported in literature. 

\subsection{Propagation in a uniform tube}
\label{sec:oneUniformTube}
In this subsection, we compare the numerical results with analytical ones for a pulse wave on a single uniform vessel ($\partial_x (\beta\sqrt{A_0})=\partial_x \beta=0$). 
To avoid reflections, nonreflecting B.C. is set at the outflow to mimic a semi-infinite tube.
Adding a small perturbation ($(\epsilon \tilde{A} , \epsilon \tilde{Q})$) to the equilibrium solution ($U=(A_0, 0)$), substituting it into the  governing
equations and dropping the terms with quadratics of
$\epsilon$, we obtain the equations for the perturbations in a linear form:
\begin{equation} 
\frac{\partial \tilde{A}}{\partial t}+\frac{\partial
\tilde{Q}}{\partial x}=0 , \quad \frac{\partial \tilde{Q}}{\partial
t}+ c_0^2\frac{\partial \tilde{A}}{\partial x} =-\frac{C_f}{A_0}
\tilde{Q}+C_v \frac{\partial ^2 \tilde{Q}}{\partial x^2} 
\label{linearized}
\end{equation} 
with $c_0=\sqrt{\frac{\beta}{2\rho} \sqrt{A_0}}$, the Moens-Korteweg celerity.
To investigate the propagation phenomena at first, we drop the non-homogeneous part ($C_f=0$ and $C_v=0$).
Then Eqs.~(\ref{linearized}) become d'Alembert equations, which admit the pure wave solution.
We assume that the initial condition is at equilibrium and 
the inflow is prescribed as $Q(0,t)=Q_{in}(t)$ with
\[
Q_{in}(t)=Q_c\sin(\frac{2\pi}{T_c}t)H(-t+\frac{T_c}{2}), \quad t>0,
\] 
where $H(t)$ is the Heaviside function, $T_c$ the period of the sinusoidal wave and $Q_c$ the amplitude. 
The solution is $c_0\tilde{A}= \tilde{Q}=Q_{in}(x -c_0t)$,
which means that the waveform propagates to the right with a speed of $c_0$.

\begin{figure}
\subfigure[]{\label{fig:con_time1}\includegraphics[width=0.5\textwidth]{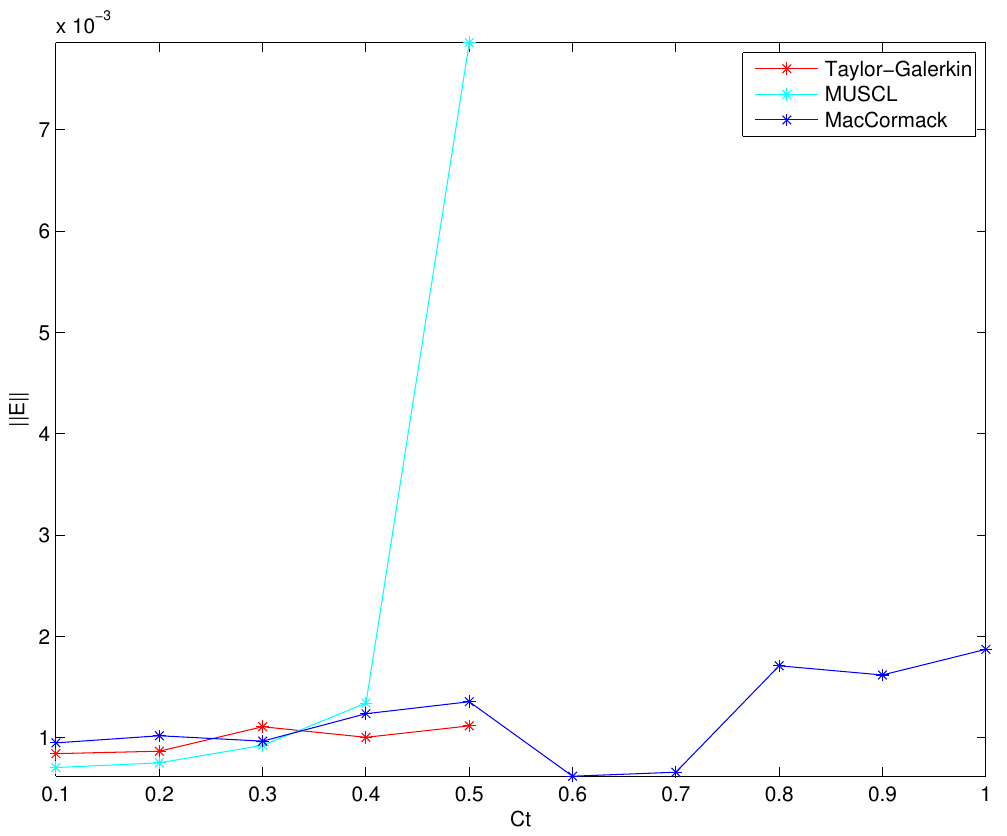}   } 
\subfigure[]{\label{fig:con_time2}\includegraphics[width=0.5\textwidth]{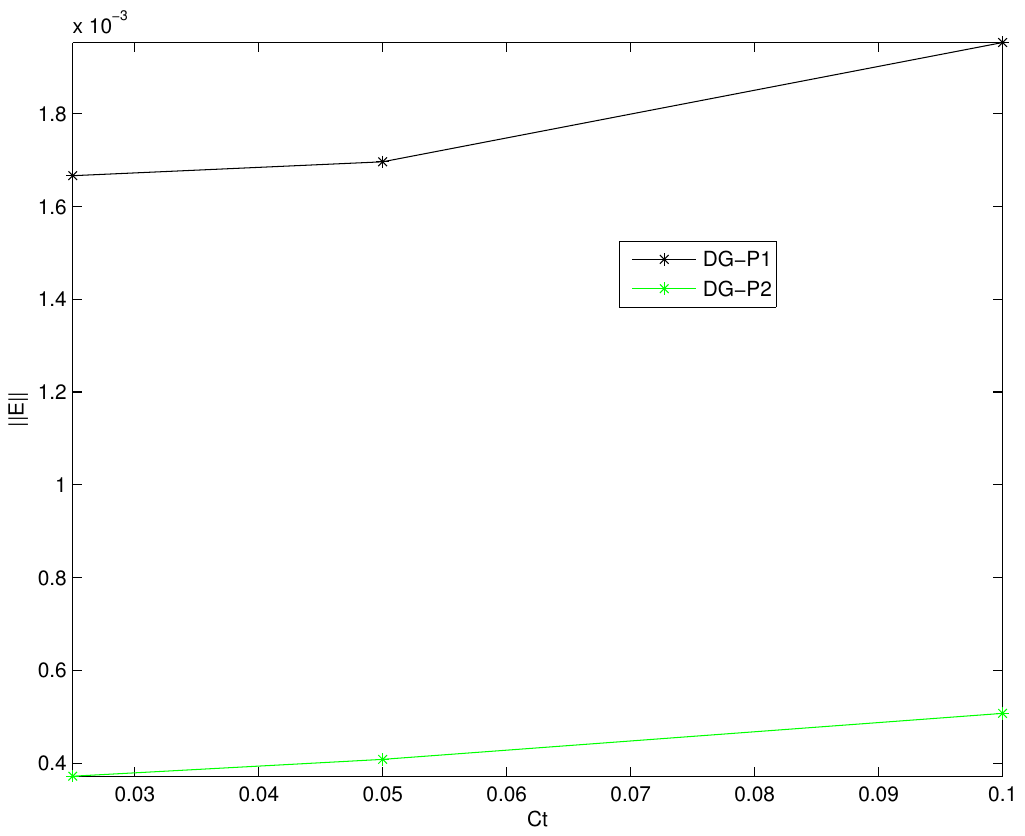} }\\
\subfigure[]{\label{fig:space_con}\includegraphics[width=0.5\textwidth]{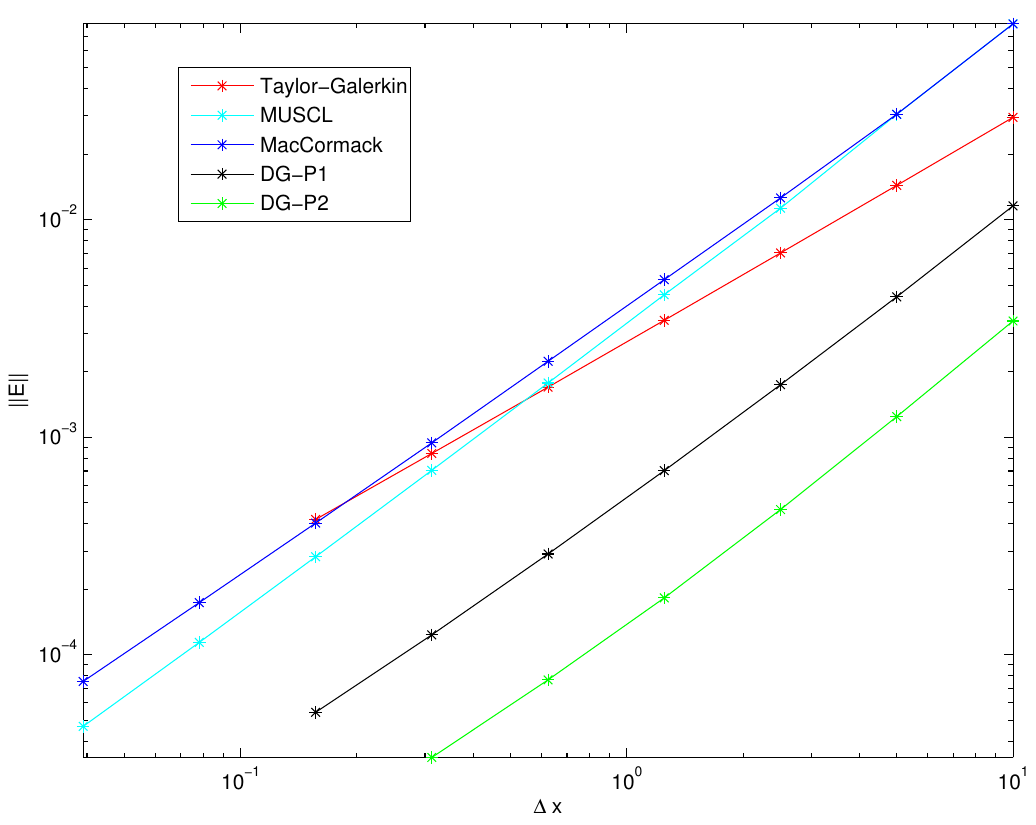}}
\subfigure[]{\label{fig:convection}\includegraphics[width=0.5\textwidth]{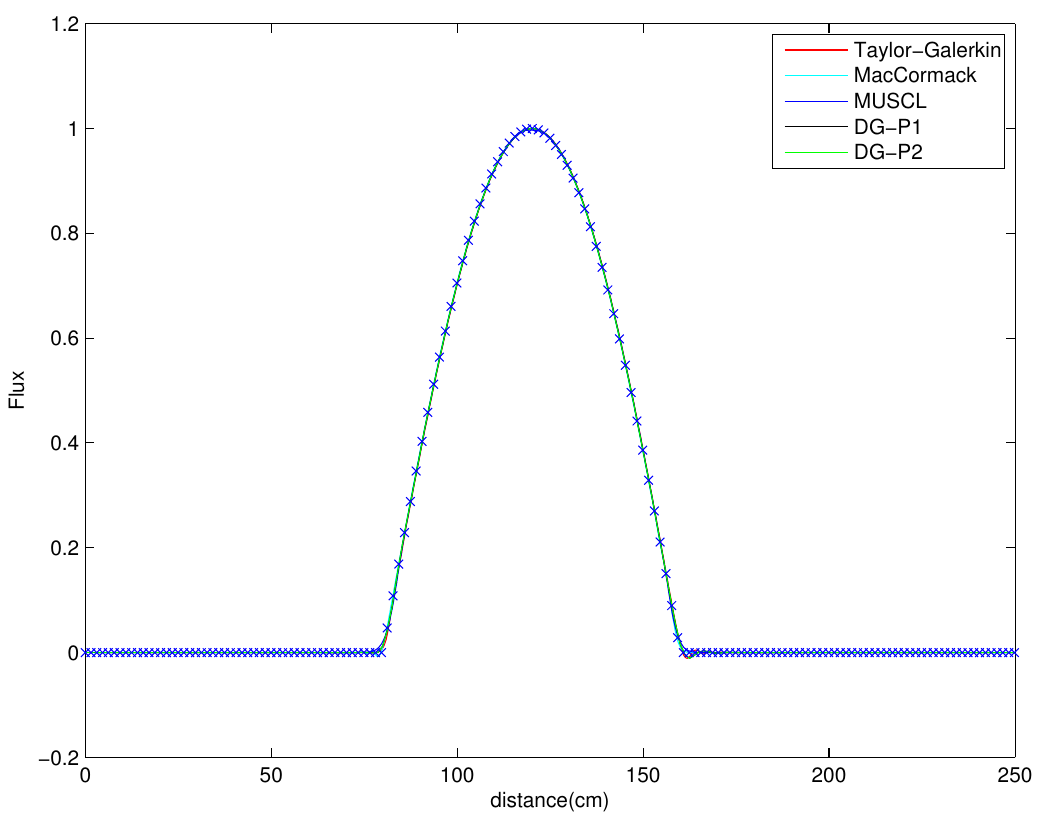} }\\
\subfigure[]{\label{fig:time_consumed}\includegraphics[width=0.5\textwidth]{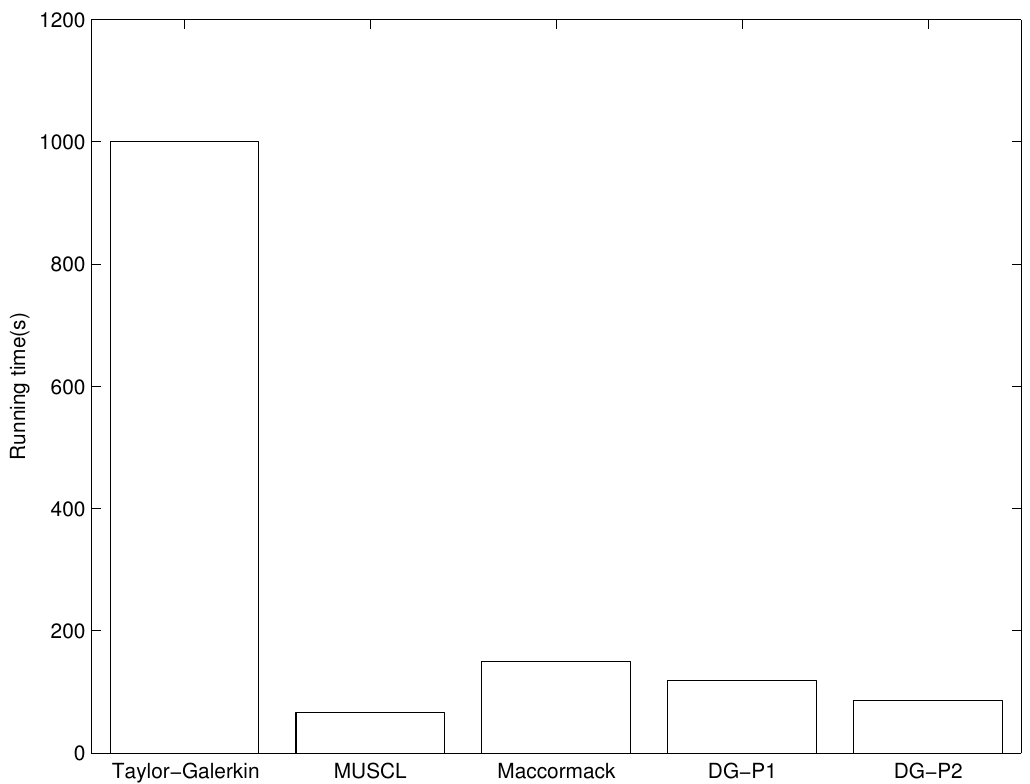} }
\subfigure[]{\label{fig:error_conSpe}\includegraphics[width=0.5\textwidth]{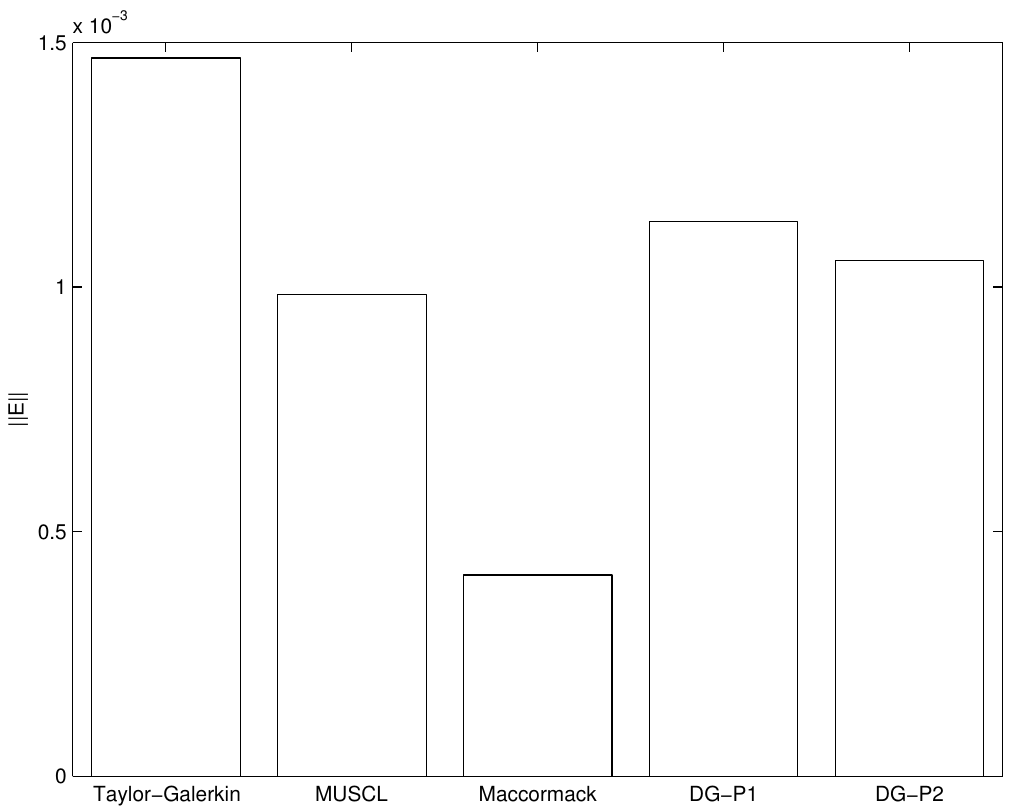} }

\caption{Test on a uniform tube. 
Top left and right: With a fixed mesh ($N_{TG}=N_{MUSCL}=N_{FD}=800, N_{DG-\mathcal{P}_1}=N_{DG-\mathcal{P}_2}=100$), errors as functions of coefficient $C_t$.
Middle left: Errors as functions of the sizes of elements (cells). 
Middle right: All the numerical solutions for the pulse wave at time 0.4s are overlapped, the analytical solution is indicated by cross signs. 
Bottom left and right: Running time and error of each scheme for the configuration shown in Table \ref{tab:meshtimeStep}. }
\end{figure}

We propose a numerical test with parameters 
of the tube inspired by~\cite{sherwin2003computational}: $L=250$cm, $A_0=3.2168 \text{cm}^2$, $ \beta=1.8734\times 10^{6}\text{Pa/m}$,
$\rho=1.050 \times 10^{3}\text{kg/m}^3$, and accordingly $c_0=400\text{cm/s}$.
To impose a small perturbation, we choose $Q_c=1\text{ml/s}$ and $T_c=0.4\text{s}$.
In this case the change ratio of the radius is $\Delta R/R_0=Q_c/(2 A_0 c_0)=0.04 \%$, thus the perturbation is assured small enough.
We take the linearized analytical solution at time $t=0.4s$ as reference to compute the errors of  the numerical solutions. 
The normalized error is defined by $||E||=||\mathbf{Q}_{numerical}-\mathbf{Q}_{analytical}||_{rms}/Q_c$, where $|| \cdot ||_{rms}$ stands for the root-mean-square error.
To specify the time step, we note that it first should satisfy the CFL (Courant-Friedrichs-Lewy) condition which writes
\begin{equation*}
\Delta t \leqslant  n_{CFL} \min_{i=0}^{N+1} \Bigl[ \frac{h_i}{\max( \frac{Q_i}{A_i}+c_i,\frac{Q_{i+1}}{A_{i+1}}+c_{i+1} )} \Bigr],	
\end{equation*}
where $h_i$ is the element (cell) size. 
For the second order Taylor-Galerkin scheme, a linear stability analysis shows that $n_{CFL}=\frac{\sqrt{3}}{3}$~\cite{formaggia2003one}.
For the second order MUSCL, $n_{CFL}=\frac{1}{2}$~\cite{delestre2012well}. 
Practice shows that $n_{CFL}=1$ for MacCormack scheme~\cite{elad1991numerical}.
A sharp estimation of the coefficient $n_{CFL}$ for the DG scheme is challenging.
We define an approximate formula, $\Delta t=C_t\frac{L}{Nc_0}$, to test the stability.
In our test, the approximate threshold values of $C_t$ for the schemes to become unstable are: 0.5 for MUSCL, 0.56 for Taylor-Galerkin and 1.0 for MacCormack.
The results agree with the report in literature. 
For the DG scheme, the time step formula is modified accordingly as
$\Delta t=\frac{C_t}{\mathcal{P}}\frac{L}{Nc_0}$,
with $\mathcal{P}$ the degree of the polynomial.
For the DG scheme, $C_t$ can not be greater than 0.1 (see Figure~\ref{fig:con_time2}). 

To further test the temporal convergence, we fix the mesh
($N_{TG}=N_{FV}=N_{FD}=800$, $N_{DG-\mathcal{P}_1}=N_{DG-\mathcal{P}_2}=100$)
and plot the numerical errors
as a function of $C_t$ (see Figure~\ref{fig:con_time1}). 
The errors vary slightly for all of the schemes except MUSCL. 
For the convergence of the temporal integration, the MUSCL scheme has to 
choose a smaller time step than  the value prescribed by the CFL condition. 
But note this is only a test in linear case, in practical applications, the coefficient $C_t$ may be much smaller for convergence (Section~\ref{seca}). 

To test the spatial convergence, we fix $C_t=0.1$, and vary the number of mesh nodes $N$.
The log-log plot of $||E||$ against $\Delta x$ can be seen in Figure~\ref{fig:space_con}.  
We have two main observations. First,  all of the schemes converge with an order between 1 and 2 and the DG
scheme converges faster (see Figure~\ref{fig:space_con}).
Second,  as shown by Figure~\ref{fig:convection} the differences between the analytical 
solution and all of the numerical solutions are hardly discernible with a moderate number of mesh points ($N_{TG}=N_{FV}=N_{FD}=800$, $N_{DG-\mathcal{P}_1}=N_{DG-\mathcal{P}_2}=100$).

To compare the actual speed  and accuracy of the four schemes,
we set $N$ and $C_t$ (see Table~\ref{tab:meshtimeStep}) such that the errors achieve the same order of magnitude (see Figure \ref{fig:error_conSpe}).
Except the Taylor-Galerkin scheme, all the schemes have the similar accuracy with very close running time (see Figure~\ref{fig:time_consumed} and~\ref{fig:error_conSpe}).
At this point, the Taylor-Galerkin scheme shows the worst accuracy and needs to run the longest time.
We note that large global matrices arise in Taylor-Galerkin scheme while the operators in other schemes are local and have small size.
That explains the relative poor performance of Taylor-Galerkin even though a larger time step is allowed by this scheme.     
We will see that in case of a network of real size, the largest number of $N$ is about 100 and Taylor-Galerkin shows a good balanced property between accuracy and speed (Section~\ref{seca}).

\begin{table}[!ht]
\centering
\begin{tabular}{c c c}
scheme & N & $C_t$   \\
\hline
Taylor-Galerkin & 800 & 0.5   \\
MUSCL     & 800 & 0.3  \\
MacCormack &1600  & 0.5 \\
DG-$\mathcal{P}_1$ &200 &0.1 \\
DG-$\mathcal{P}_2$ &100 & 0.1
\end{tabular}
\caption{Number of elements and coefficient of time step}
\label{tab:meshtimeStep} 
\end{table}

\subsection{Attenuation due to the viscosity of blood}
We now consider the same linearized Eq.~(\ref{linearized}) with the small 
 term due to skin friction ($C_f \ne 0$ and $C_v=0$).
The main dynamics of the system 
will be grossly the same traveling wave but attenuated by viscosity of blood.
This  behaviour can be predicted by asymptotic analysis. 
We have a small non-dimensional parameter $\epsilon_f=T_cC_f/A_0$, 
which is the ratio of the characteristic time of pulse $T_c$ to the characteristic time of attenuation $A_0/C_f$. 
 In order to see how the waveform
slowly evolves when it propagates to, say right, we make a change of
variables to $\tau=\epsilon_f t$ and $\xi=x-c_0t$ (slow time, moving frame).
The two differential operators $\partial _t$ and $\partial _x$ expand as
\begin{align*} \frac{\partial}{ \partial t}=\frac{\partial
\tau}{\partial {t}}\frac{\partial}{\partial \tau}+\frac{\partial \xi }{\partial
t} \frac{\partial}{\partial \xi}=\epsilon_f \frac{\partial}{\partial \tau}-c_0
\frac{\partial}{\partial \xi} \\ \frac{\partial}{\partial
x}=\frac{\partial \xi}{\partial x}\frac{\partial}{\partial \xi}
=\frac{\partial}{\partial \xi}.
\end{align*} 
The solution has the asymptotic expansion
\begin{equation*} \tilde {A}=\tilde {A}_0+\epsilon_f \tilde {A}_1+...,
\quad \tilde {Q}=\tilde {Q}_0+\epsilon_f \tilde {Q}_1+...
\end{equation*} 
Substituting these into the governing equations
expressed in new variables and collecting the terms with the same
order of $\epsilon_f$, one has
\begin{align*} (-c_0\frac{\partial \tilde{A}_0} {\partial
\xi}+\frac{\partial \tilde{Q}_0}{\partial \xi})+\epsilon_f(
\frac{\partial \tilde{A}_0}{\partial \tau}-c_0\frac{\partial
\tilde{A}_1}{\partial \xi}+\frac{\partial \tilde{Q}_1}{\partial \xi})
+..=0\\ (-c_0\frac{\partial \tilde{Q}_0}{\partial \xi}+c_0^2
\frac{\partial \tilde{A}_0}{\partial \xi})+\epsilon_f( \frac{\partial
\tilde{Q}_0}{\partial \tau}-c_0\frac{\partial \tilde{Q}_1}{\partial
\xi} +c_0^2 \frac{\partial \tilde{A}_1}{\partial \xi}+ \frac{\tilde{Q}_0}{T_c})
+..=0.
\end{align*} 
We take the first order term in $\epsilon_f$ in the first equation, substitute it in 
the first order term in $\epsilon_f$ in the second equation. 
Then we obtain
$$
( \frac{\partial
\tilde{Q}_0}{\partial \tau}+c_0
\frac{\partial \tilde{A}_0}{\partial \tau} +  \frac{\tilde{Q}_0}{T_c})=0.
$$
From the terms of the zeroth
order in $\epsilon_f$, which involve derivative in $\xi$ only, the solution must have the form
$\tilde{Q}_0=c_0\tilde{A}_0(\tau,\xi)+\delta(\tau)$.
Substituting it into the previous equation generates terms $\frac{\partial \delta}{\partial \tau} $ and $\delta(\tau)$. 
These are secular terms and thus can be set null.
So we have $c_0\tilde{A}_0=\tilde{Q}_0$ and
$\frac{\partial \tilde{Q}_0}{\partial \tau}=-\frac{1}{2T_c}\tilde{Q}_0$
, or
\[
\tilde{Q}_0=\tilde{Q}_0(0,\xi)e^{-\tau/(2T_c)}=\tilde{Q}_0(0,x-c_0t)e^{-\epsilon_f
t /(2T_c)}.
\]
For more on asymptotic analysis of blood flow in large blood vessels, we refer to reference~\cite{yomosa1987solitary}.

\begin{figure}[ht]
\centering
\includegraphics[width=0.5\textwidth]{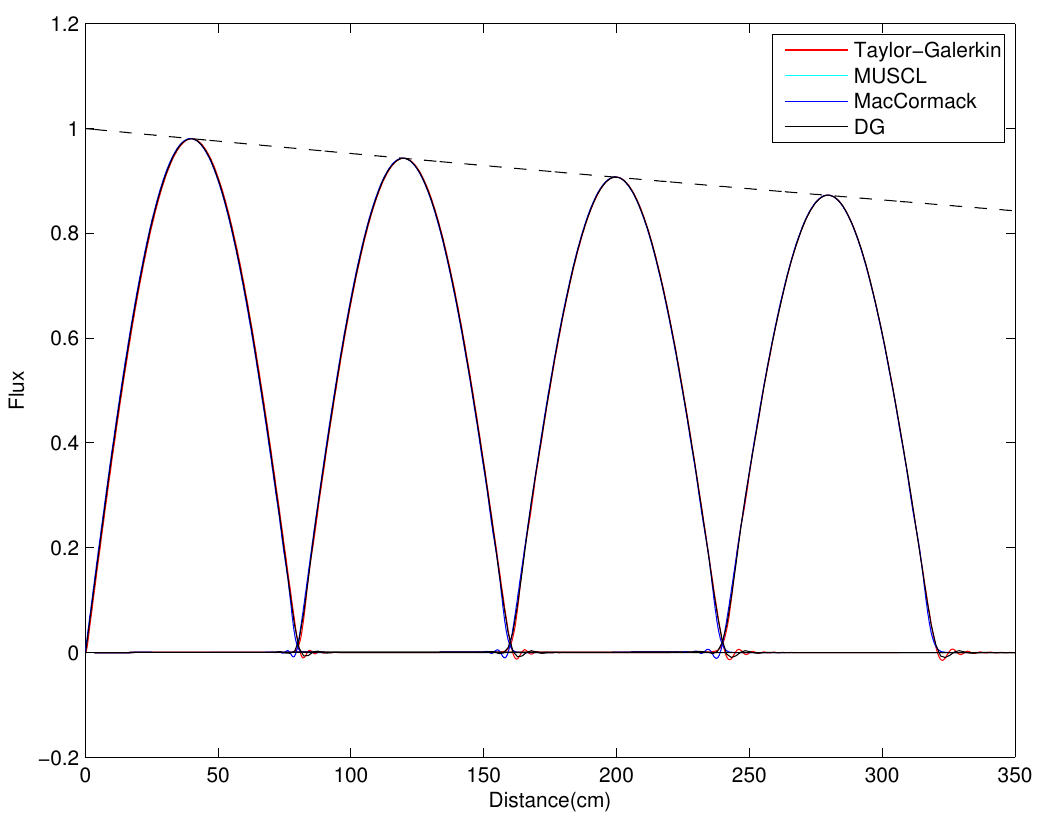}
\caption{Attenuation due to the skin friction. The snapshots are at time
0.2s, 0.4s, 0.6s and 0.8s.
The dashed line is $\exp({-\frac{C_fx}{2A_0c_0}})$ with 
$2A_0c_0/C_f \simeq 2000\text{cm}$.
The flux is normalized with respect to $Q_c$.}
\label{Attenuation}
\end{figure}

In Figure~\ref{Attenuation}, we plot the snapshots of the waveform at time 0.2s, 0.4s,
0.6s and 0.8s. In the computation, the inflow is a half sinusoidal flux as described in the previous subsection and the outflow is nonreflecting.
The skin friction coefficient $C_f$ is $40 \nu \pi$, and the parameter $2A_0c_0/C_f$ is about $2000\text{cm}$.
The damping rate of the amplitude of the waveform agrees very well with the analytical prediction,
$\exp({-\frac{C_fx}{2A_0c_0}})$, which is indicated by the dashed line. 
Also note that the errors of different schemes are not the same.
The MUSCL scheme causes the peak of the wave to slightly flatten, while all of the other schemes are dispersive: we 
have small oscillations at the foot of the signal.

\subsection{Diffusion due to the viscosity of the arterial wall}
\label{sec:resultDiffusion}
\begin{figure}[ht]
\centering
\includegraphics[width=0.5\textwidth]{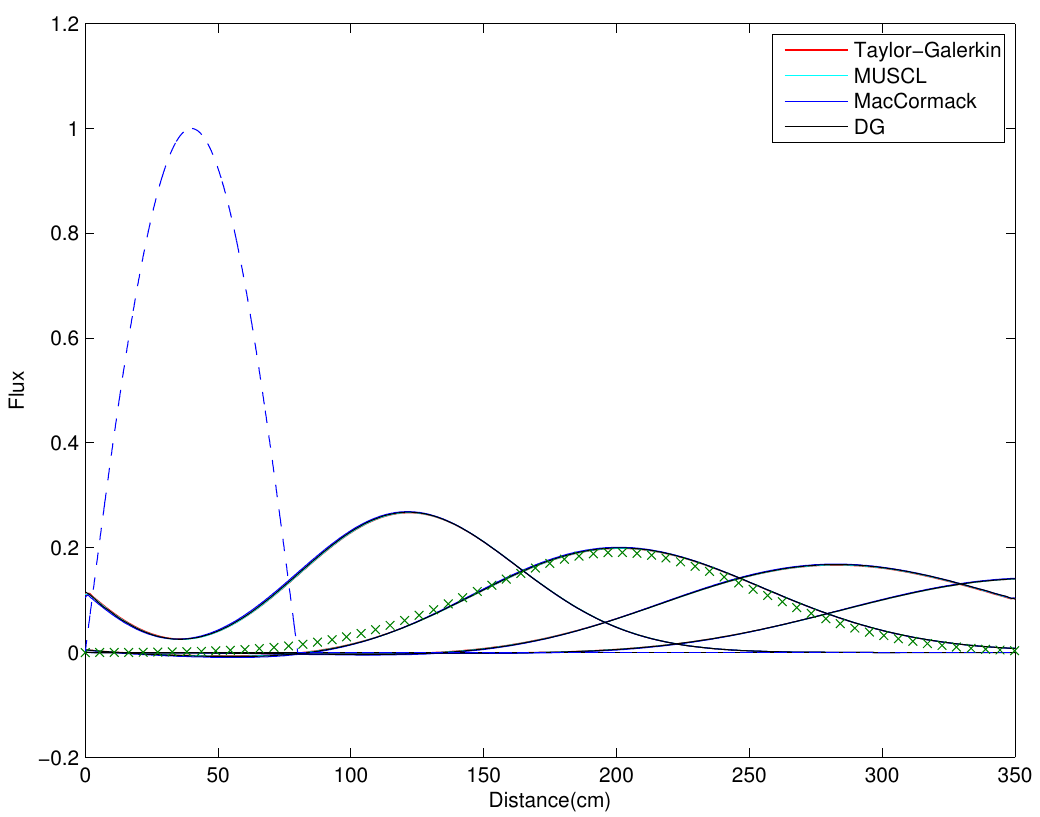}
\caption{Diffusion due to the viscosity of the wall.
The dashed line is the initial condition.
One half of the original waveform propagates
to right. The snapshots are at time
0.2s, 0.4s, 0.6s and 0.8s.
The analytical prediction from the convolution at time 0.4s is indicated by cross signs. 
The difference between the different numerical solutions is not discernible.
The flux is normalized with respect to $Q_c$.}
\label{diffusion}
\end{figure}

This time we consider the linearized Eqs.~(\ref{linearized}) with the Kelvin-Voigt effect but no viscous fluid effect ($C_f=0$ and $C_v \ne 0$).
The small parameter is now  $\epsilon_v=C_v/(c_0^2T_c)$.
If we apply the same technique as described in the previous subsection, we can readily obtain
the diffusive  behaviour of the pulse wave in the moving frame:
\begin{equation} \frac{\partial \tilde{Q}_0}{\partial
\tau}=\frac{c_0^2T_c}{2}\frac{\partial^2 \tilde{Q}_0}{\partial^2 \xi}.
\label {heatEq}
\end{equation} 
The solution of this equation can be given by the
convolution
\[ \tilde{Q}_0(\tau,\xi)=\int^{+\infty}_{-\infty} \tilde{Q}_0(0,\xi)
G(\tau,\xi-\zeta) d\zeta
\] where $G$ is the fundamental solution of the Eq.~(\ref{heatEq})
\[ G(\tau,\xi)=\frac{1}{\sqrt{2\pi\tau c_0^2T_c}}e^{-\xi ^2/(2\tau c_0^2T_c)}
\]
and $\tilde{Q}_0(0,\xi)$ is the initial state.
In the test vessel, the parameters are kept the same as in the case
of attenuation. The coefficient $C_v$ is 0.6275$\text{m}^2/\text{s}$ and $\epsilon_v\simeq 0.1$. This corresponds to 
$\phi$=5000Pa $\cdot$ s, which is in the range of observed values on animals~\cite{armentano1995arterial}.
To facilitate the calculation
of the analytical solution, nonreflecting B.C.s are imposed at the two ends of the
vessel and the I.C. is a half sinusoidal waveform for $Q$ (dashed line in Figure~\ref{diffusion}) 
and a constant value for $A_0$.
It is clear that half of the initial wave propagates 
to right and at the same time the waveform is spread out due to the diffusive effect.
The analytical solution at time 0.4s (indicated by cross signs) agrees well
with the corresponding numerical solutions.  

Another point worthy noticing is the operator splitting errors. In the DG scheme,
no operator splitting error is induced. All of the other numerical schemes
adopt operator splitting method. They produce very accurate solutions as well as DG.
Thus it verifies the \textit{a priori} judgement that Godunov splitting is sufficient.

\subsection{Shock-like phenomena due to the nonlinearity}

We now consider the full nonlinear system,
but without any source terms ($C_f=0$ and $C_v = 0$).
The small parameter is now  $\epsilon_2=Q_c/(c_0 A_0)$.
If we apply the same
technique as described in the previous subsection, we can readily obtain
an equation for the nonlinear  behaviour of the pulse wave in the moving frame (inviscid Burgers' equation):
\begin{equation*} \frac{\partial \tilde{Q}_0}{\partial
\tau}=\frac{1}{2A_0} \tilde{Q}_0 \frac{\partial \tilde{Q}_0}{\partial \xi}
\label {nonlin}
\end{equation*}

\begin{figure}[h]
\addtolength{\tabcolsep}{-2pt}
\subfigure[]{\label{fig:non}\includegraphics[width=0.5\textwidth]{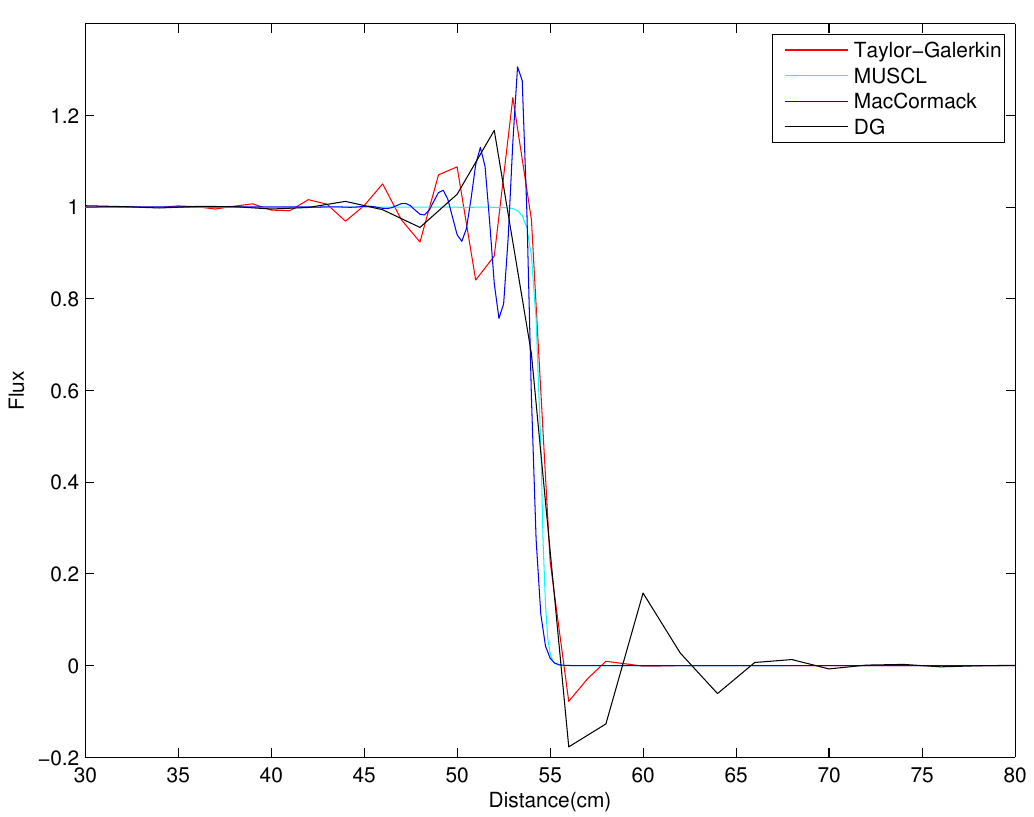} }
\subfigure[]{\label{fig:con_visco7}\includegraphics[width=0.5\textwidth]{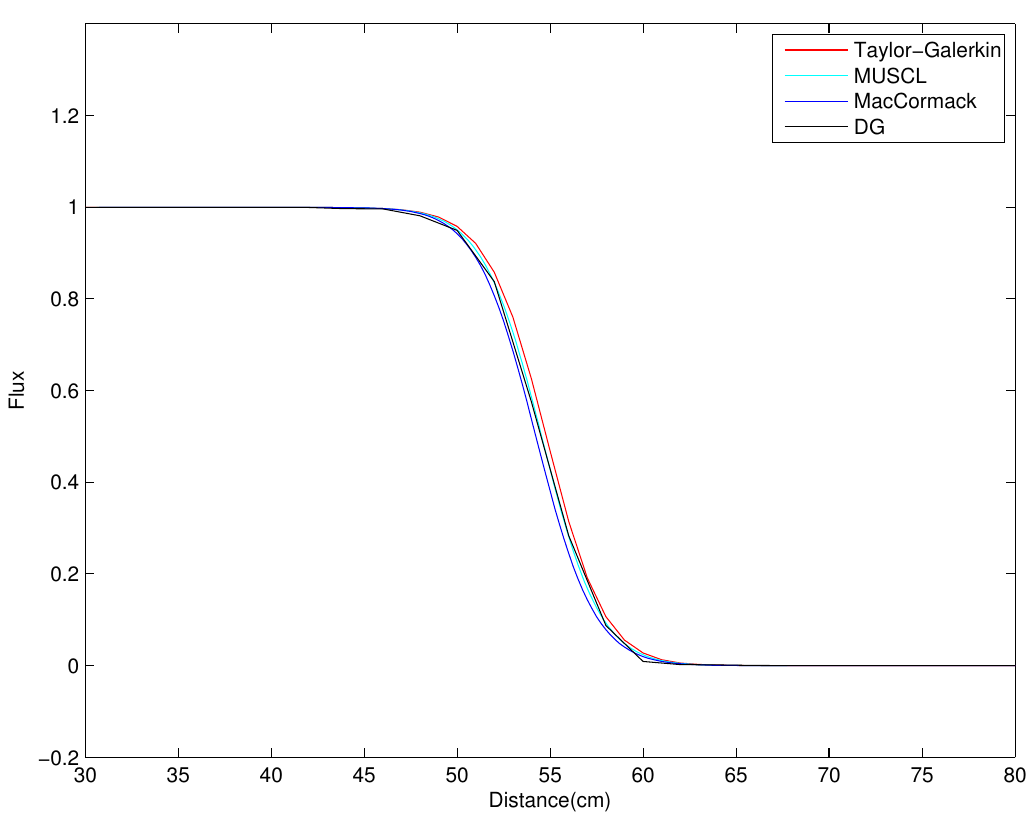}   } \\
\caption{A shock in the system. A step jump signal of flux is imposed at the inlet and a snapshot is shown.
The left figure (a) shows that the MUSCL scheme with a flux limiter captures the shock without nonphysical oscillations, whereas the other numerical schemes cause spurious oscillations. 
The right figure (b) shows that all the schemes give almost the same result for a system with a moderate
physical diffusive term.}
\label{nonlinearity}
\end{figure}

One important consequence of nonlinear hyperbolic system is that shocks may arise even
if the initial condition is very smooth.
In normal physiological conditions, shocks are not observed in arterial systems.
But in venous system, shock-like phenomena may occur on muscular veins during
walking and running. The intramuscular pressure (equivalent to
$P_{ext} $ in our model) can rise to $20-40$ kPa in a few milliseconds~\cite{ballard1998leg}.
In such situation, experiments and numerical
simulations~\cite{flaud2012experiments,marchandise2010accurate} have shown this critical
behaviour. 
For some large mammals, for instance giraffes, even in static postures, the gravity-driven flow in a long inclined vein
may develop into shock-like waves,
like the roll waves in a shallow-water channel~\cite{brook1999numerical,brook2002model}. 
For another example, the traumatic rupture of the aorta is responsible for a significant
percentage of traffic death and the rupture may be well accounted for by the shock-like transition resulted from the
blunt impact to the thorax~\cite{kivity1974nonlinear}.  
For possible applications in these situations, we test all the schemes with a shock-like wave.

To generate a shock, we impose a step jump signal of flux at the inlet.
For a vessel of 1 meter, the numbers of elements for Taylor-Galerkin, MacCormack and MUSCL schemes are 100, 200 and 800 respectively. 
The DG scheme uses 25 elements and the order of polynomial is 2.  
Figure~\ref{nonlinearity} shows that the MUSCL scheme with a flux limiter captures the shock without nonphysical oscillations,
whereas the other numerical schemes cause spurious oscillations. 
This verifies the total-variation-diminishing (TVD) property of the MUSCL scheme.  
But the MUSCL is very diffusive at the shock, thus a very fine mesh is required.
For the DG scheme, limiters may be introduced as well to eliminate the oscillations~\cite{hesthaven2008nodal}.
This remedy will be necessary for DG to be applicable on problems with shocks.
On Figure~\ref{fig:con_visco7} we plot a case with some viscosity of the wall. 
The added moderate physical diffusive term smoothens the wave and all the schemes give almost
the same result.

\subsection{Reflection and transmission at a branching point}

Up to now, we focused on the various behaviours of wave within a single vessel: propagation, attenuation, diffusion, etc.
Now, we look at the boundaries of each artery. 
Indeed, pressure waves are reflected and transmitted at the conjunction points of a network.
For a linearized system, given the impedance $Z=\frac{\rho c_0}{A_0}$, the reflection and transmission coefficients at a branching point
can be calculated by the formula,
\begin{equation}
\mathcal{R}=\frac{Z_p^{-1}-(Z_{d_1}^{-1}+Z_{d_2}^{-1})}{Z_p^{-1}+(Z_{d_1}^{-1}+Z_{d_2}^{-1})}, \quad
\mathcal{T}=\frac{2Z_p^{-1}}{Z_p^{-1}+(Z_{d_1}^{-1}+Z_{d_2}^{-1})},
\label{reflTransCoef}
\end{equation}
where $Z_p$ and $Z_d$ are the characteristic impedance of the parent and daughter vessels~\cite{fung1997biomechanics,pedleyfluid}.

\begin{figure}[ht]
\centering
\includegraphics[width=0.5\textwidth]{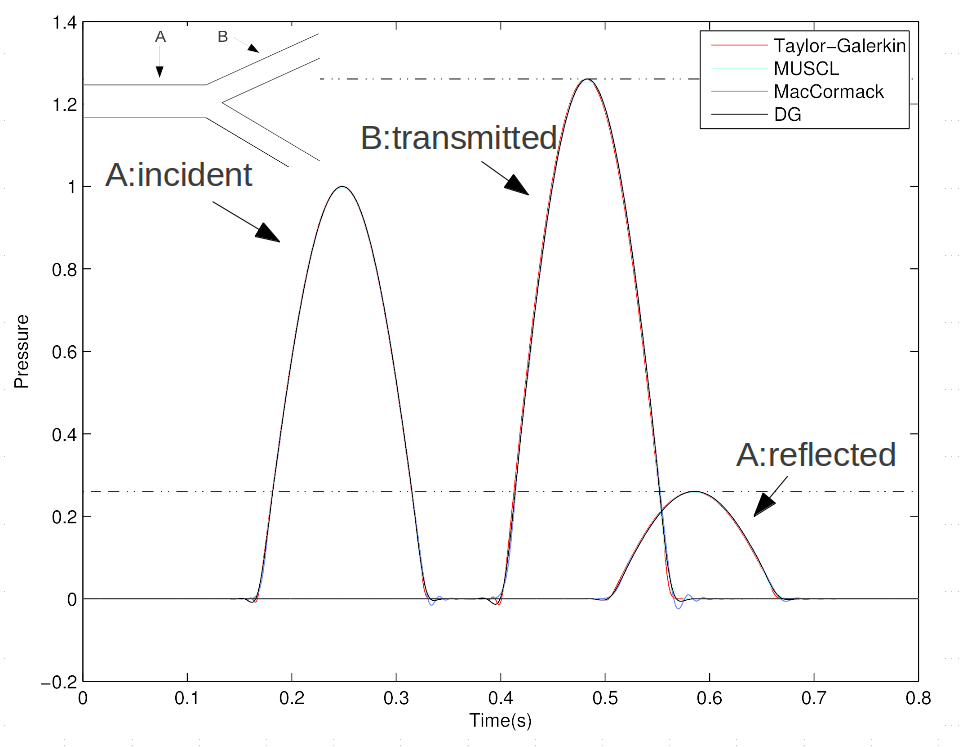}
\caption{Reflection and transmission of pressure wave at a branching point.
The time profiles of pressure at points A and B are plotted.
The analytical reflection and transmission coefficients are 0.2603 and 1.2603
(indicated by the dashed line).}
\label{BranchingPressure}
\end{figure}

In Figure~\ref{BranchingPressure}, 
for sake of illustration, the configuration of the branching and the time profiles of pressure 
at two locations are shown. The amplitude is normalized with respect to $Q_c=1\times 10^{-6} \text{m}^3\text{/s}=1\text{ml/s}$.
For the parent vessel: $\beta=2.3633\times 10^6 \text{Pa}/\text{m}$, $A_0=4\text{cm}^2$ and for each of the daughter vessels:
$\beta=6.3021\times 10^6  \text{Pa/m}$, $A_0=1.5\text{cm}^2$.
The B.C.s at the outlets of the daughter vessels are nonreflecting. 
Thus the reflected pulse wave is generated at the conjunction point.
According to the formula~(\ref{reflTransCoef}), $\mathcal{R}=0.2603$ and $\mathcal{T}=1.2603$. 
The pressure profiles at the points A and B agree very well with the analytical predictions. 
All of the numerical schemes are compatible with this treatment of conjunction point.
Note that in healthy arterial system, the related arteries of most conjunctions are well matched
such that there are essentially no reflections ($\mathcal{R}$=0) at the conjunctions~\cite{wang2004wave,papageorgiou1990area}.
The purpose of the proposed configuration is just to test the numerical schemes.

\subsection{Application on a full systematic arterial network}
\label{seca}

As already mentioned in the introduction, a relatively realistic description of arterial system has been done in 
1D simulations, with different numerical solvers by different teams.
For example, in~\cite{mynard20081d,sherwin2003computational}, Galerkin approach is used.
In these papers, wall viscosity is not included. 
Note that ~\cite{reymond2009validation} gives a survey of literature on the details the model,
and adopted a viscoelastic model of the wall.
But, in all of those papers,  usually only one numerical scheme is adopted and cross comparisons among them are not available.
In this subsection, we compute a network of 55 arteries with the viscoelastic model presented above and make a cross comparison
among the numerical schemes. 
To this end, the topology and properties value of the arterial network are adapted from~\cite{sherwin2003computational}.
But the viscosity coefficient of the Kelvin-Voigt model on human body is not given in this paper. 
In reference~\cite{armentano1995arterial}, the viscosity of aortic wall of dogs was modeled by a Kelvin-Voigt model and
it shows that the value of $\phi$ is in the range of $3.8\pm1.3 \times 10^3 \text{Pa}\cdot \text{s}$
to $7.8\pm1.1\times 10^3 \text{Pa}\cdot \text{s}$.
Hence, we assume $\phi=5\times 10^3 \text{Pa}\cdot \text{s}$ to calculate the coefficient $C_v$. 
The final parameters of the network we used are shown in Table~\ref{tab:systemic_network}.
We note that there may be differences between arteries in human and dog and the 
arteries in different locations may cause a considerable variation.
Nevertheless the inclusion of viscosity term makes it possible to
test the numerical schemes in a more realistic condition.
\begin{table}[ht!]
\centering
\begin{threeparttable}
\caption{Arterial network\tnote{}} \label{tab:systemic_network}
\centering \scriptsize
\begin{tabular}{c c c c c c c} \hline\hline 
  &  &  $l$ & $A_0$  & $\beta$ & $C_v$  & \\
ID & Name & $(\text{cm})$ & $(\text{cm}^2)$ & $(10^6 \text{Pa/cm})$& $(10^4 \text{cm}^2\text{/s})$ & $R_t$ \\ \hline
1 & Ascending aorta & 4.0 & 6.789 & 0.023 & 0.352 & --\\
2 & Aortic arch I & 2.0 & 5.011 & 0.024   & 0.317 & -- \\
3 & Brachiocephalic & 3.4 & 1.535 & 0.049 & 0.363 & -- \\
4 & R.subclavian I & 3.4 & 0.919 & 0.069  & 0.393 & -- \\
5 & R.carotid & 17.7 & 0.703 & 0.085      & 0.423 & -- \\
6 & R.vertebral & 14.8 & 0.181 & 0.470    & 0.595 & 0.906 \\
7 & R. subclavian II & 42.2 & 0.833 & 0.076 & 0.413 & -- \\
8 & R.radius & 23.5 & 0.423 & 0.192       & 0.372 & 0.82 \\
9 & R.ulnar I & 6.7 & 0.648 & 0.134       & 0.322 & -- \\
10 & R.interosseous & 7.9 & 0.118 & 0.895 & 0.458 & 0.956 \\
11 & R.ulnar II & 17.1 & 0.589 & 0.148    & 0.337 & 0.893\\
12 & R.int.carotid & 17.6 & 0.458 & 0.186 & 0.374 & 0.784 \\
13 & R. ext. carotid & 17.7 & 0.458 & 0.173 & 0.349 & 0.79 \\
14 & Aortic arch II & 3.9 & 4.486 & 0.024 & 0.306 & -- \\
15 & L. carotid & 20.8 & 0.536 & 0.111    & 0.484 & -- \\ 
16 & L. int. carotid & 17.6 & 0.350 & 0.243 & 0.428 & 0.784 \\
17 & L. ext. carotid & 17.7 & 0.350 & 0.227 & 0.399 & 0.791 \\
18 & Thoracic aorta I & 5.2 & 3.941 & 0.026 & 0.312 & --\\
19 & L. subclavian I & 3.4 & 0.706 & 0.088  & 0.442 & -- \\
20 & L. vertebral & 14.8 & 0.129 & 0.657    & 0.704 & 0.906 \\
21 & L. subclavian II & 42.2 & 0.650 & 0.097 & 0.467 & -- \\
22 & L. radius & 23.5 & 0.330 & 0.247       & 0.421 & 0.821 \\
23 & L. ulnar I & 6.7 & 0.505 & 0.172       & 0.364 & -- \\
24 & L. interosseous & 7.9 & 0.093 & 1.139  & 0.517 & 0.956 \\
25 & L. ulnar II & 17.1 & 0.461 & 0.189     & 0.381 & 0.893 \\
26 & intercostals & 8.0 & 0.316 & 0.147    & 0.491 & 0.627 \\
27 & Thoracic aorta II & 10.4 & 3.604 & 0.026 & 0.296 & -- \\
28 & Abdominal aorta I & 5.3 & 2.659 & 0.032 & 0.311 & -- \\
29 & Celiac I & 2.0 & 1.086 & 0.056        & 0.346 & -- \\
30 & Celiac II & 1.0 & 0.126 & 0.481       & 1.016 & -- \\
31 & Hepatic & 6.6 & 0.659 & 0.070         & 0.340 & 0.925 \\
32 & Gastric & 7.1 & 0.442 & 0.096         & 0.381 & 0.921 \\
33 & Splenic & 6.3 & 0.468 & 0.109         & 0.444 & 0.93 \\
34 & Sup. mesenteric & 5.9 & 0.782 & 0.083 & 0.439 & 0.934 \\
35 & Abdominal aorta II & 1.0 & 2.233 & 0.034 & 0.301 & -- \\
36 & L. renal & 3.2 & 0.385 & 0.130        & 0.481 & 0.861 \\ 
37 & Abdominal aorta III & 1.0 & 1.981 & 0.038 & 0.320 & -- \\
38 & R. renal & 3.2 & 0.385 & 0.130        & 0.481 & 0.861 \\ 
39 & Abdominal aorta IV & 10.6 & 1.389 & 0.051 & 0.358 & -- \\ 
40 & Inf. mesenteric & 5.0 & 0.118 & 0.344  & 0.704 & 0.918 \\
41 & Abdominal aorta V & 1.0 & 1.251 & 0.049  & 0.327 & -- \\
42 & R. com. iliac & 5.9 & 0.694 & 0.082   & 0.405 & -- \\
43 & L. com. iliac & 5.8 & 0.694 & 0.082   & 0.405 & -- \\
44 & L. ext. iliac & 14.4 & 0.730 & 0.137  & 0.349 & -- \\
45 & L. int. iliac & 5.0 & 0.285 & 0.531   & 0.422 & 0.925 \\
46 & L. femoral & 44.3 & 0.409 & 0.231     & 0.440 & -- \\
47 & L. deep femoral & 12.6 & 0.398 & 0.223 & 0.419 & 0.885 \\
48 & L. post. tibial & 32.1 & 0.444 & 0.383 & 0.380 & 0.724 \\
49 & L. ant. tibial & 34.3 & 0.123 & 1.197  & 0.625 & 0.716 \\
50 & L. ext. iliac & 14.5 & 0.730 & 0.137   & 0.349 & -- \\ 
51 & R. int. iliac & 5.0 & 0.285 & 0.531    & 0.422 & 0.925 \\
52 & R. femoral & 44.4 & 0.409 & 0.231      & 0.440 & -- \\ 
53 & R. deep femoral & 12.7 & 0.398 & 0.223 & 0.419 & 0.888 \\
54 & R. post. tibial & 32.2 & 0.442 & 0.385 & 0.381 & 0.724 \\
55 & R. ant. tibial & 34.4 & 0.122 & 1.210  & 0.628 & 0.716 \\
\hline
\end{tabular}
\begin{tablenotes}
\item Data adapted from~\cite{armentano1995arterial} and \cite{sherwin2003computational}.
\end{tablenotes}
\end{threeparttable}
\end{table}

\begin{figure}[ht]
\addtolength{\tabcolsep}{-2pt}
\subfigure[]{\label{fig:non}\includegraphics[width=0.5\textwidth]{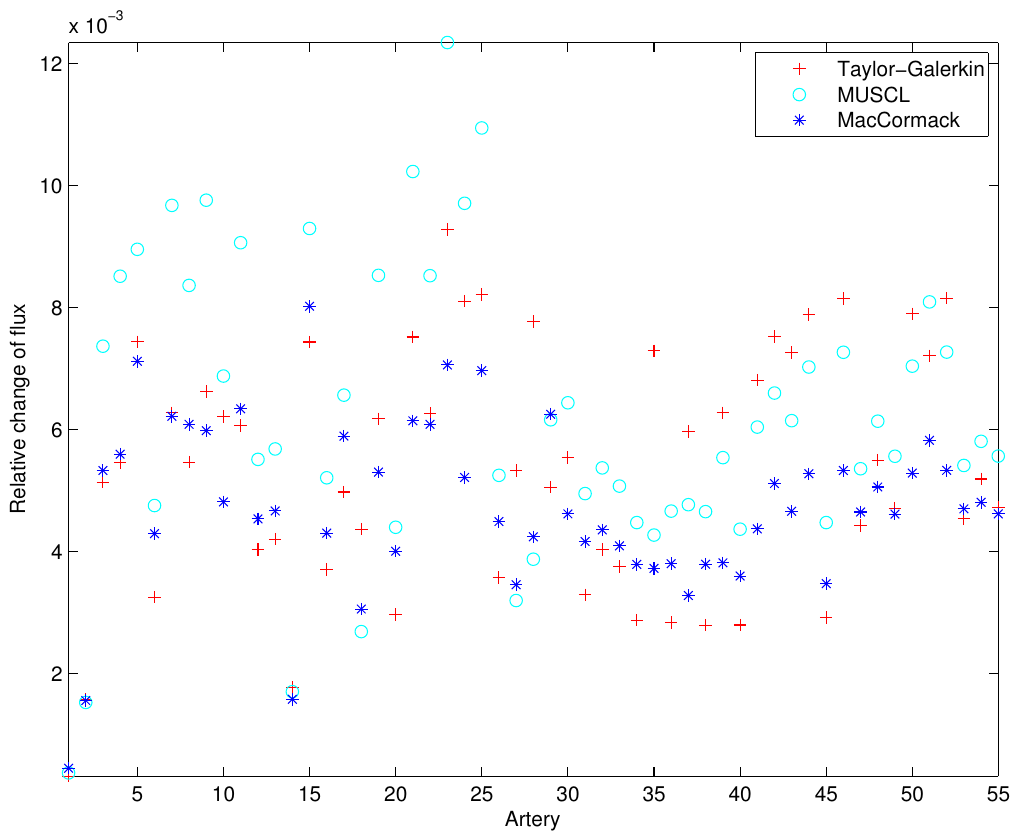} }
\subfigure[]{\label{fig:con_visco}\includegraphics[width=0.5\textwidth]{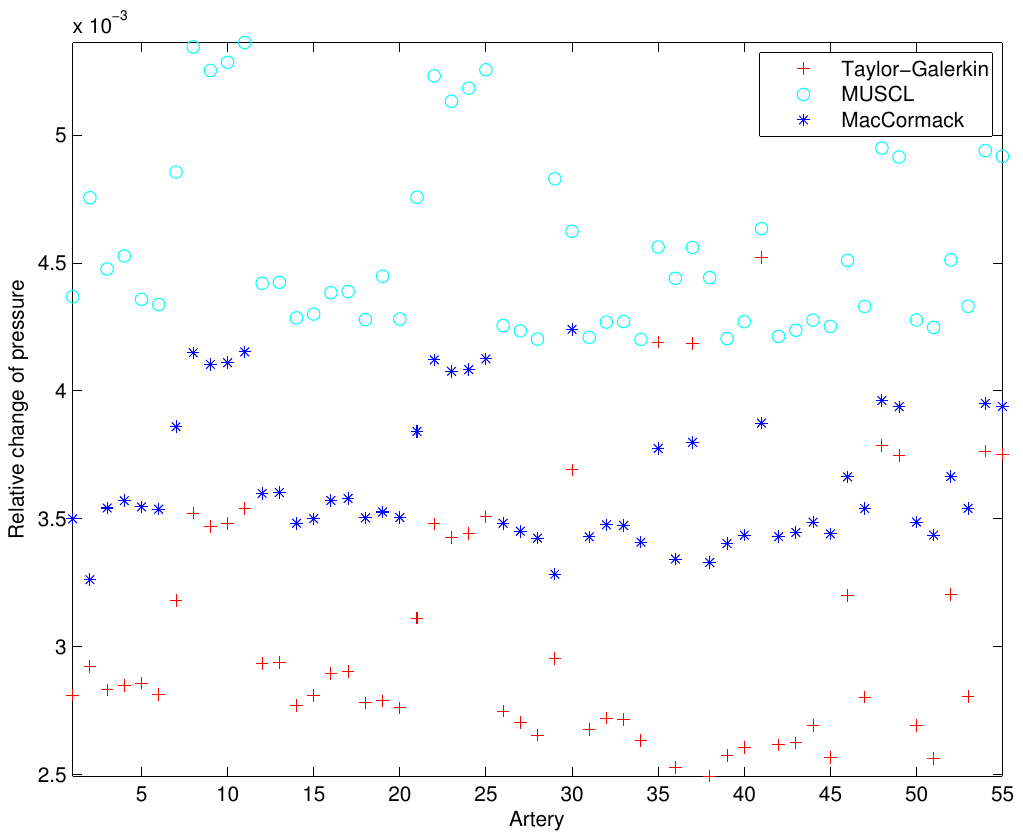}   } \\
\caption{Relative changes of the solutions when the mesh is doubled from 2$N_{base}$ to $4N_{base}$.
The left figure shows that the relative changes of all the fluxes are less than 1.3 \%.
The right figure shows that the relative changes of all the pressures are less then 0.6\% .}
\label{converge_system}
\end{figure}

\begin{figure}[ht!]
\centering
\addtolength{\tabcolsep}{-4pt}
\begin{tabular}{c c c}
& Flux & Pressure   \\
\multirow{4}{4cm}{\subfigure[]{\label{fig:wholeBody}\includegraphics[width=0.35\textwidth]{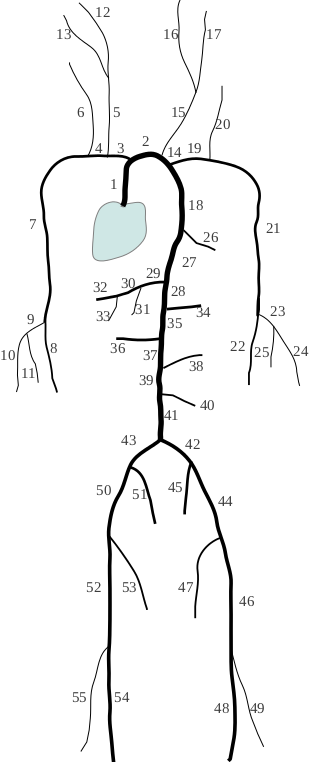}} }
& \subfigure[]{\label{fig:Flux_1}\includegraphics[width=0.3\textwidth]{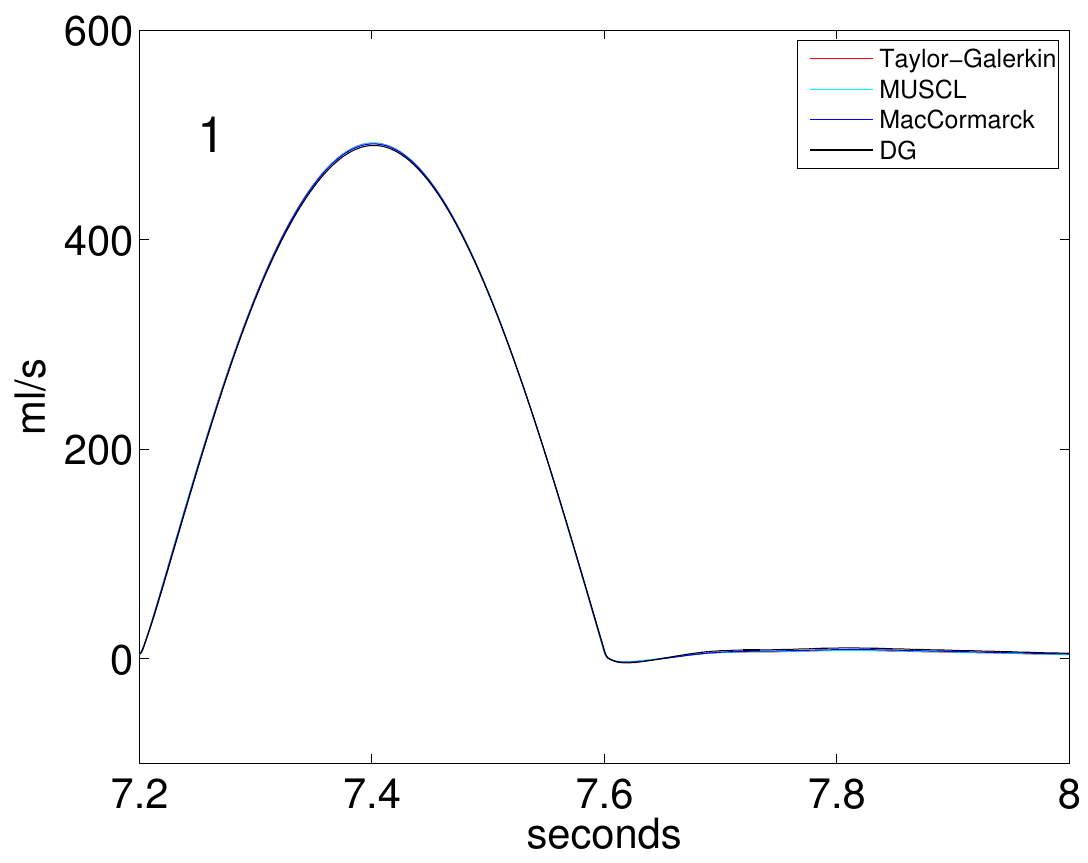} }  
& \subfigure[]{\label{fig:Pressue_1}\includegraphics[width=0.3\textwidth]{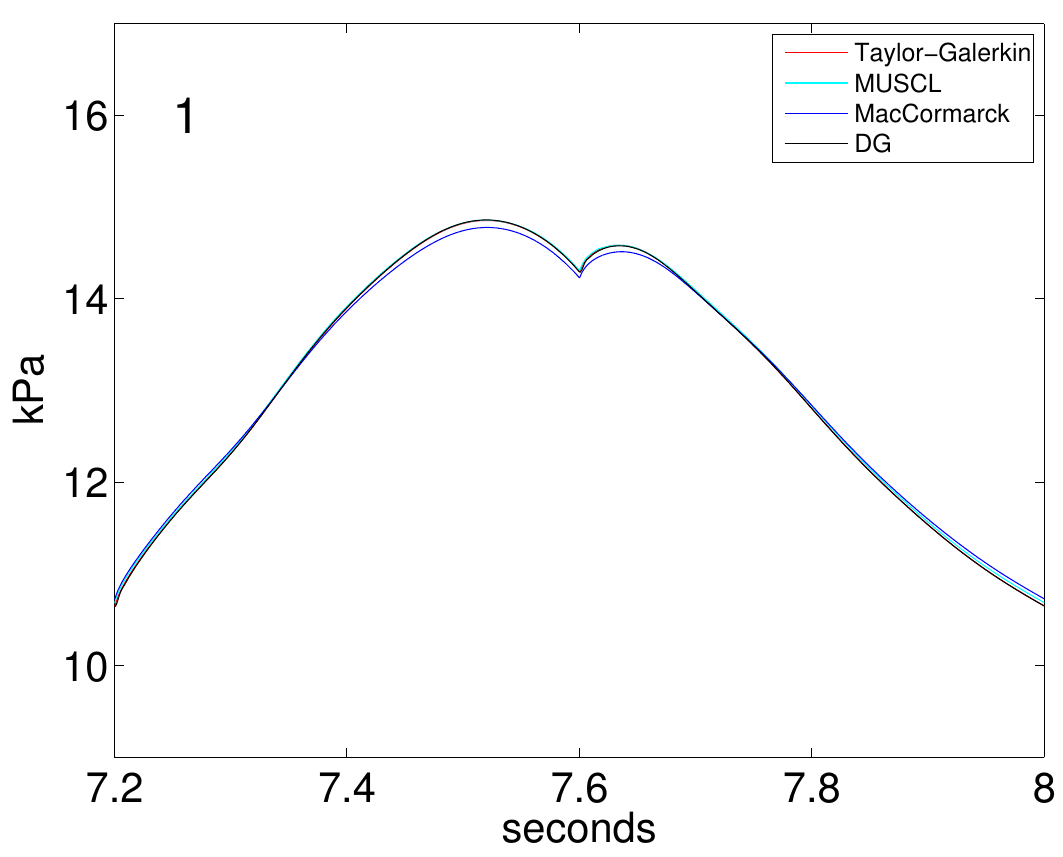} } \\

&\subfigure[]{\label{fig:Flux_8} \includegraphics[width=0.3\textwidth]{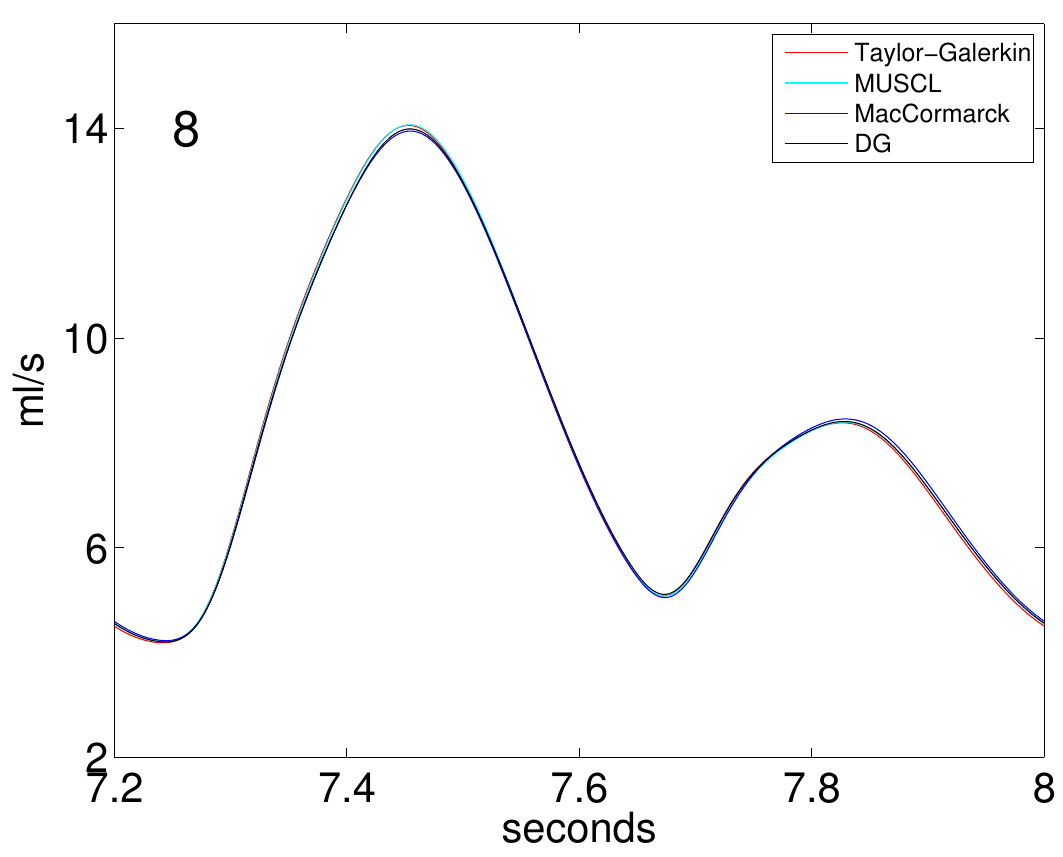} } 
&\subfigure[]{\label{fig:Pressure_8} \includegraphics[width=0.3\textwidth]{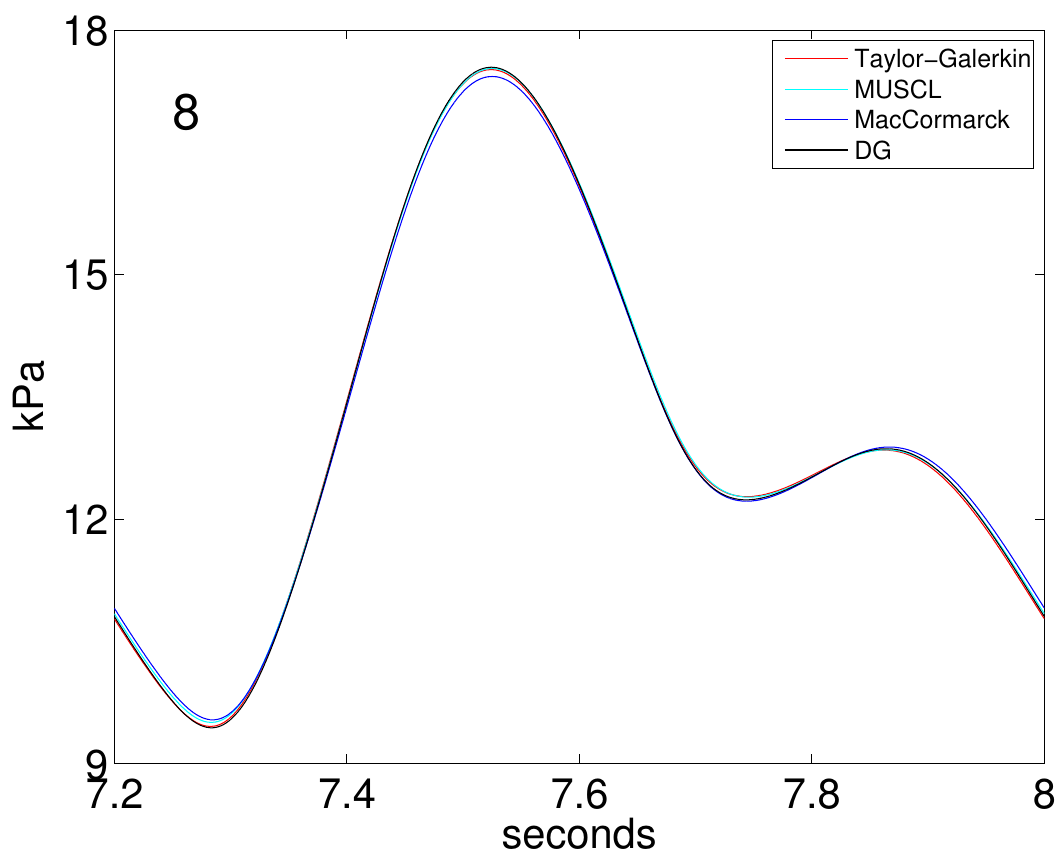} } \\

& \subfigure[]{\label{fig:Flux_37} \includegraphics[width=0.3\textwidth]{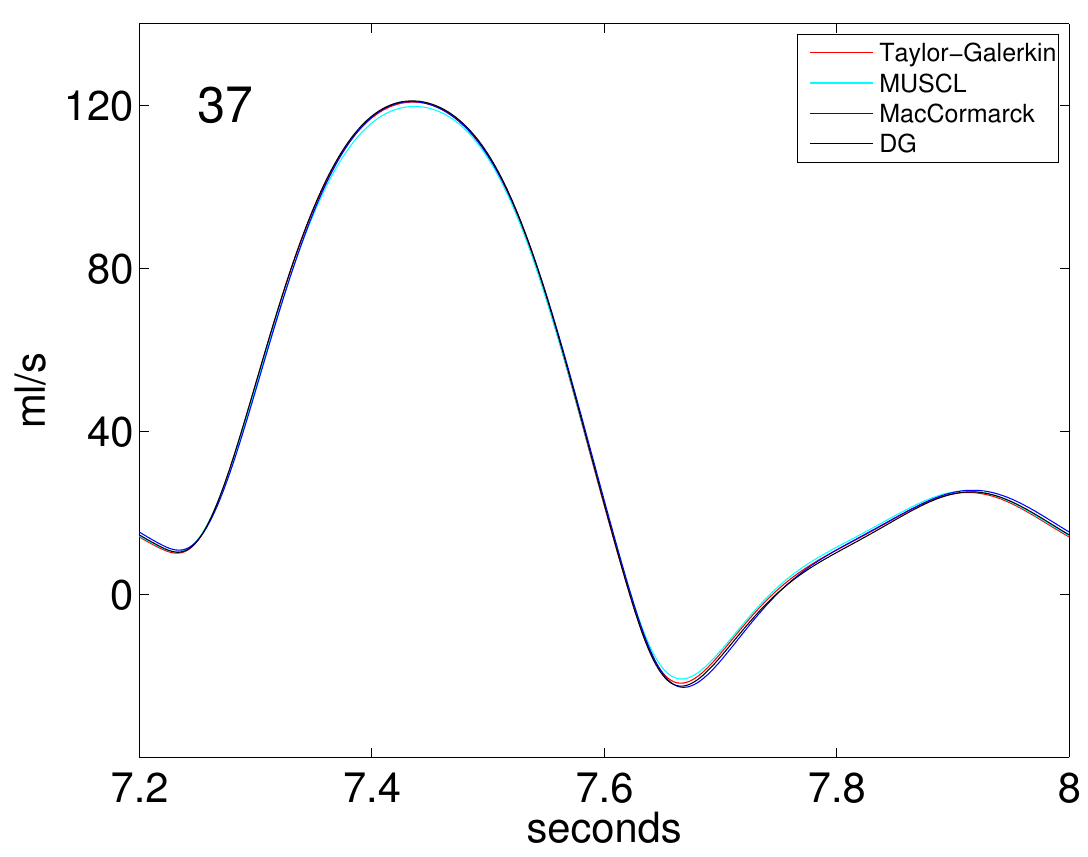} }
& \subfigure[]{\label{fig:Pressure_37} \includegraphics[width=0.3\textwidth]{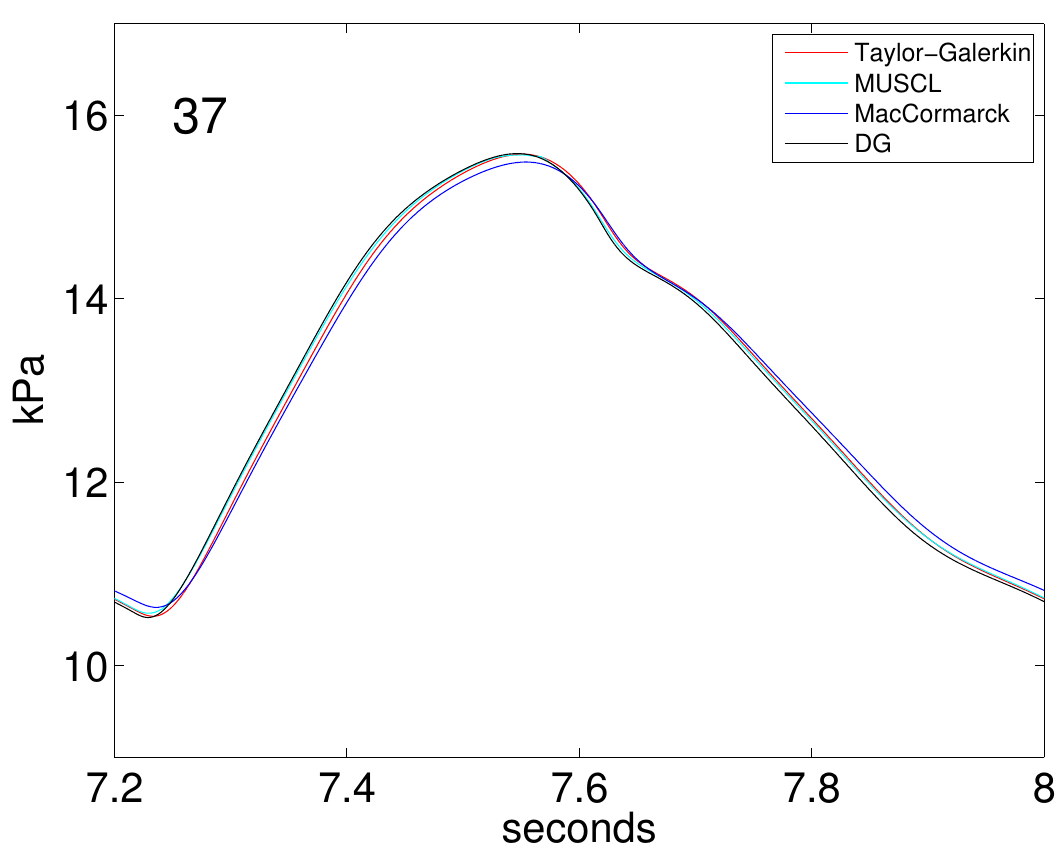} } \\

&\subfigure[]{\label{fig:Flux_54} \includegraphics[width=0.3\textwidth]{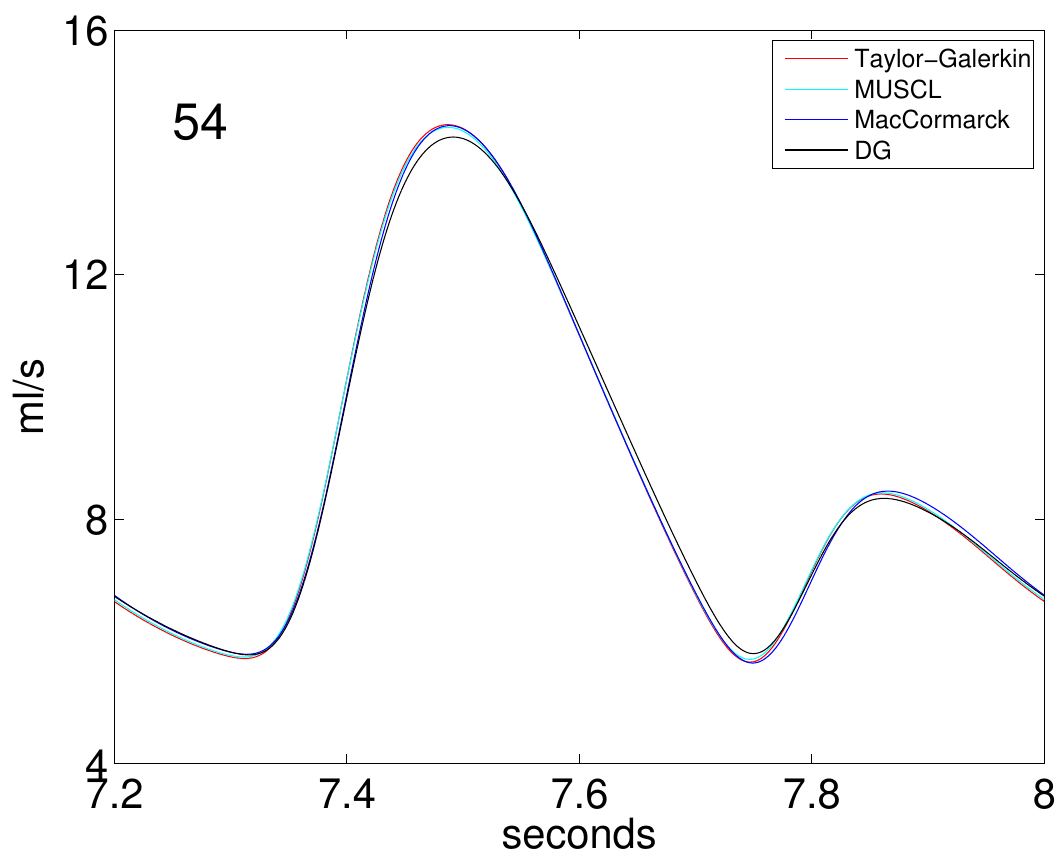} } 
&\subfigure[]{\label{fig:Pressure_54} \includegraphics[width=0.3\textwidth]{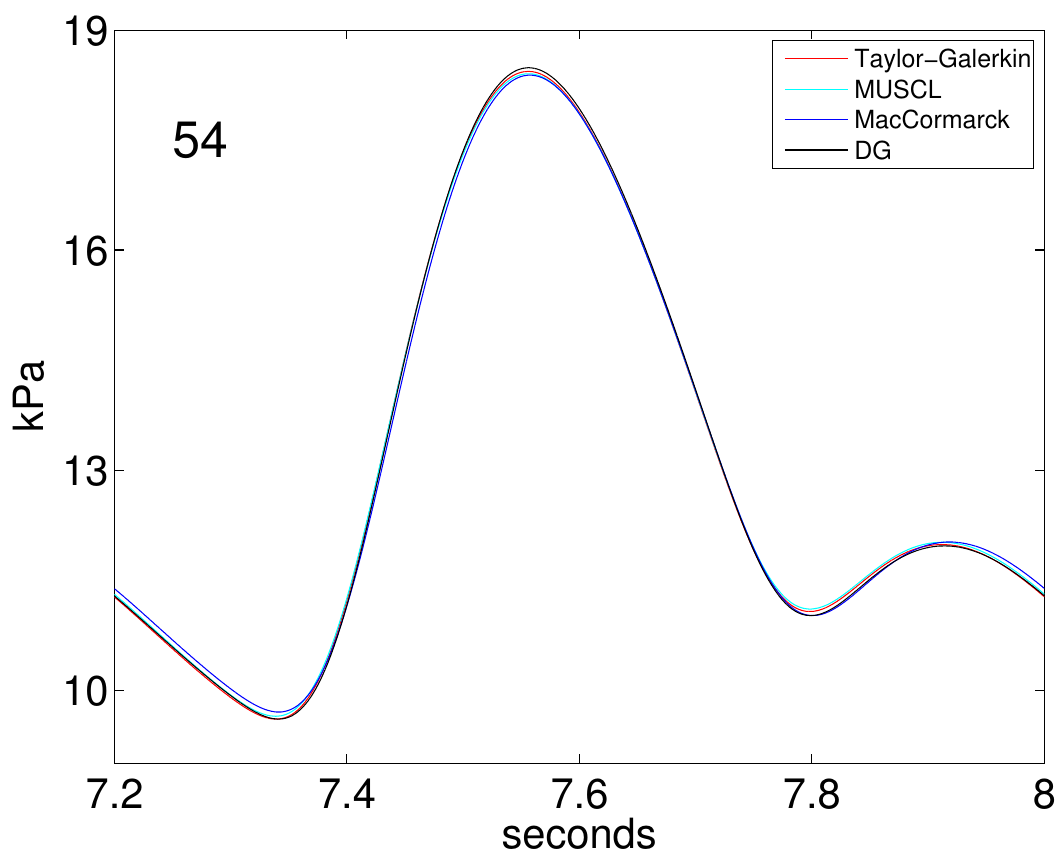} } \\

\end{tabular} 
\caption{The history profiles of flux and pressure at four locations.
Ten heart beats are computed to secure that steady state is achieved, but only the tenth heart beat is plotted.
The differences between the four numerical schemes are very small.
See Table~\ref{tab:systemic_network} for time steps and running time of each scheme.}
\label{system_profile}
\end{figure}

The peak value of the input flux $Q_c$ is 500 ml/s. 
This value is very close to the peak flow rate at the root of aortic artery~\cite{reymond2009validation}.  
We choose $\min_{i=1}^{i=55}(L^i/c_0^i)$ as a reference length,
with $L^i$ the vessel length and $c_0^i$ the linearized  wave speed of the $i$-th artery. 
For a coarsest possible mesh, the number of elements (cells) of the $i$-th artery is $N^i_{base}=\lfloor \frac{L^i/c_0^i}{\min_{i=1}^{i=55}(L^i/c_0^i)} \rfloor$,
where $\lfloor \cdot\rfloor$ is the floor function. 
We computed the relative change of solutions when the number of the elements (cells) is doubled.
Figure~\ref{converge_system} shows the relative change of the solutions 
when the number of the elements (cells) is changed from $2N_{base}$ to $4N_{base}$. 
The relative change of a quantity (for example flux $Q$) with two meshes $N_1$ and $N_2$ is defined as 
$||\mathbf{Q}_{N_1}-\mathbf{Q}_{N_2}||_{rms}/(Q_{max}-Q_{min})$, where $||\cdot||_{rms}$ is the root-mean-square error as before, $Q_{max}$ and $Q_{min}$ are the
maximum and minimum values within one heart beat. 
Figure~\ref{converge_system} shows that the changes of flux and pressure are less than 
1.5\% for all of of the schemes except DG. Thus we plotted in Figure~\ref{system_profile}
the results computed with mesh 2$N_{base}$.
The DG scheme is not tested in this manner because it is already converged:
results in Figure~\ref{system_profile} show that there is no discernible difference between the DG solutions with the others
even with the coarsest possible mesh.
In this computation, the order of polynomial of DG is 1, thus the total number of free degrees is $2N_{base}$,
which is equal to those of the other schemes. 
Time step is prescribed by $\Delta t=C_t
 \min_{i=1}^{i=55}( 
\frac{L^i}{N^i c_0^i})$. The coefficient $C_t$ and the corresponding real time steps in the computation 
are shown in Table~\ref{tab:timeSteps}.

\begin{table}[Ht]
\centering
\begin{tabular}{c c c c }
 scheme & $C_t$ & $\Delta t $ ($10^{-6}$s) &running time (min) \\
\hline
Taylor-Galerkin & 0.4 & 222 & 22.0  \\
MUSCL    & 0.25 &    139   &  31.9\\
MacCormack &0.1  & 55.5  &  91.2\\
Local DG &0.006 &   6.66 &  576\\
\end{tabular}
\caption{Time steps and running time for one heart beat using one processor on a standard Linux workstation with MATLAB.}
\label{tab:timeSteps} 
\end{table}

In Figure~\ref{system_profile} we plot the history profiles of flux and pressure at the middle of four representative arteries.
All of the numerical solutions agree very well.
The main features of the pressure and flux profiles reported in literature~\cite{sherwin2003computational,reymond2009validation}
are observed.
The peak value of pressure waveform increases as we travel down the system. We can also see the dicrotic notch at artery 1.
At artery 37, a reverse flow is observed (see Figure~\ref{fig:Flux_37}), 
which agrees with clinical measurement~\cite{reymond2009validation}.

\begin{figure}[ht]
\subfigure[]{\label{fig:Flux_elas_visco37}\includegraphics[width=0.51\textwidth]{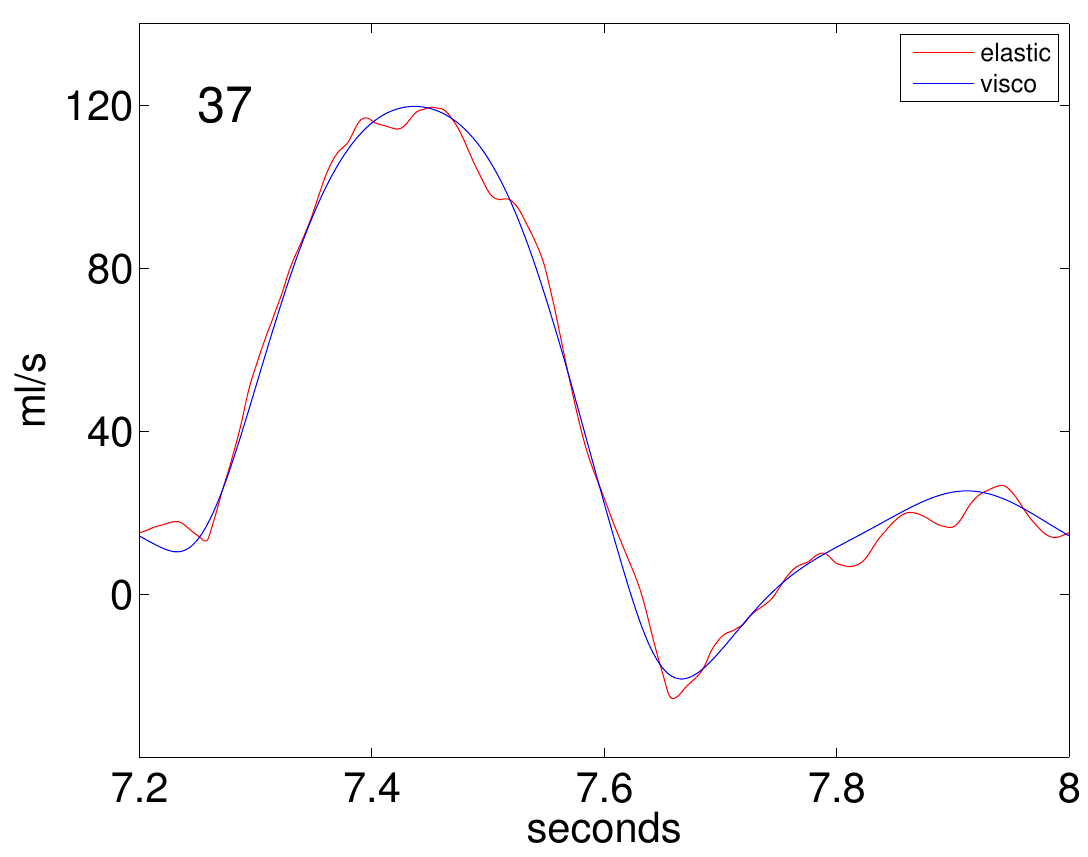}}
\subfigure[]{\label{fig:Pressure_elas_visco37}\includegraphics[width=0.5\textwidth]{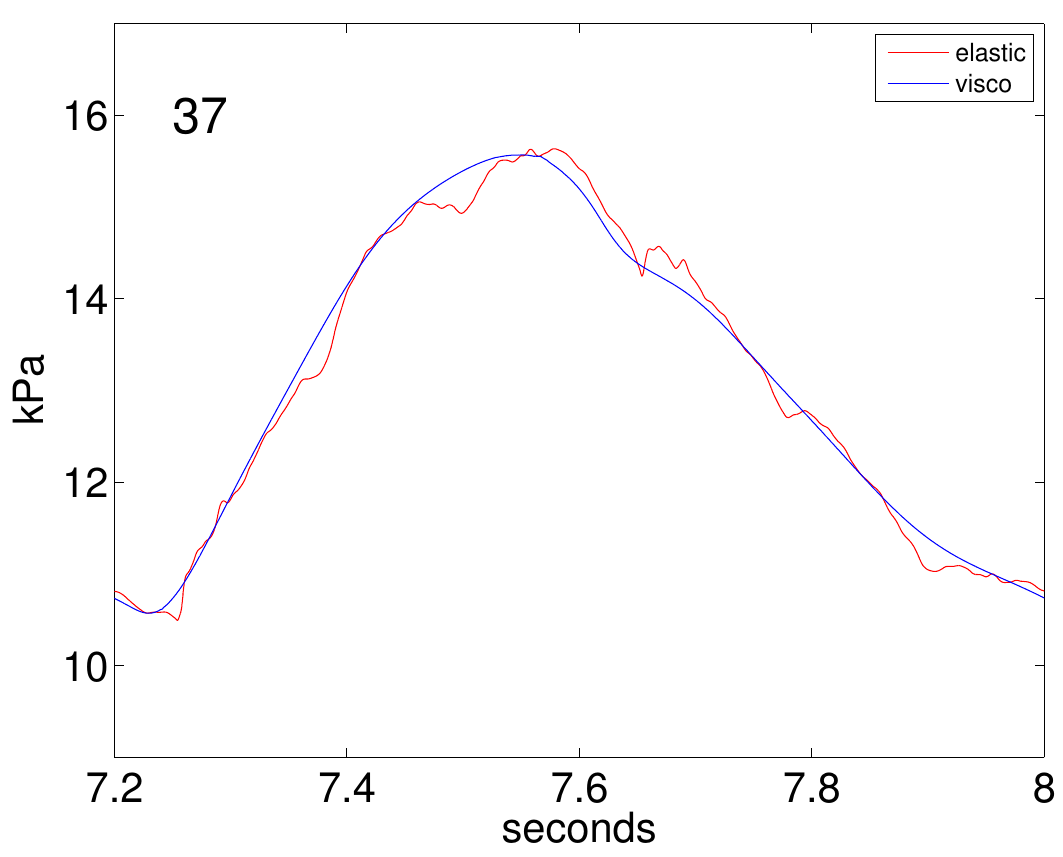} }\\
\subfigure[]{\label{fig:Flux_elas_visco54}\includegraphics[width=0.5\textwidth]{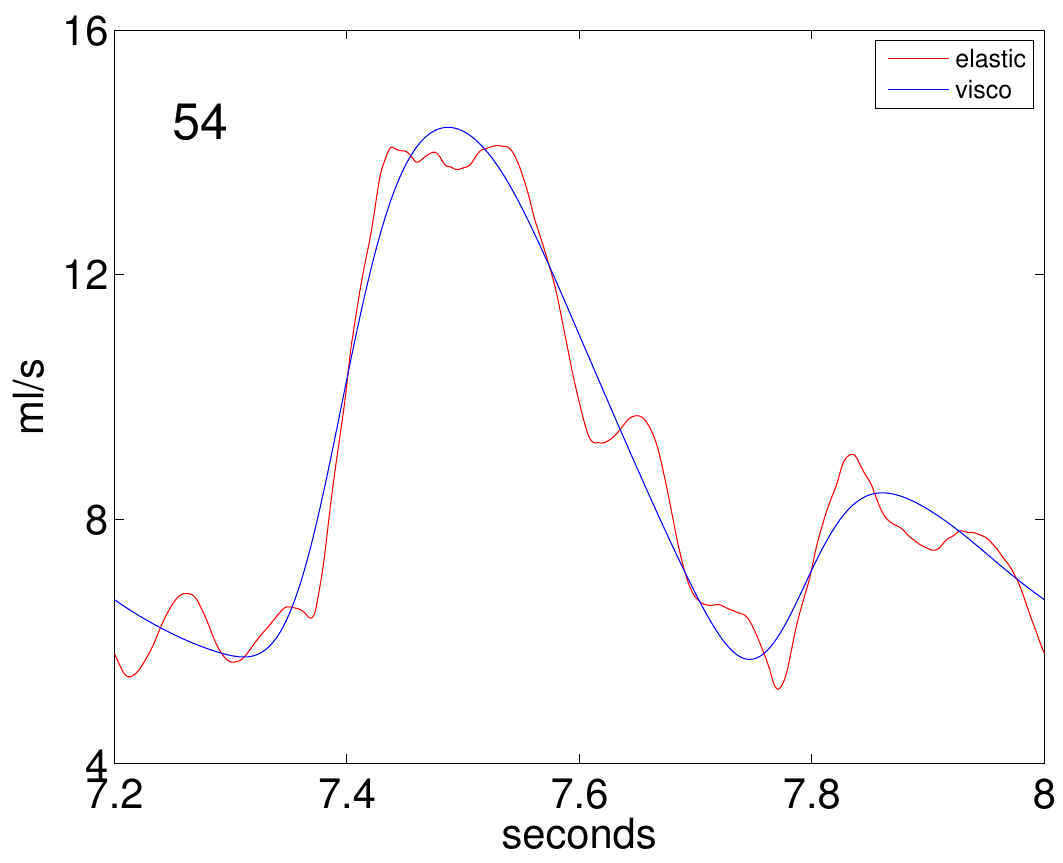}   } 
\subfigure[]{\label{fig:Pressure_elas_visco54}\includegraphics[width=0.5\textwidth]{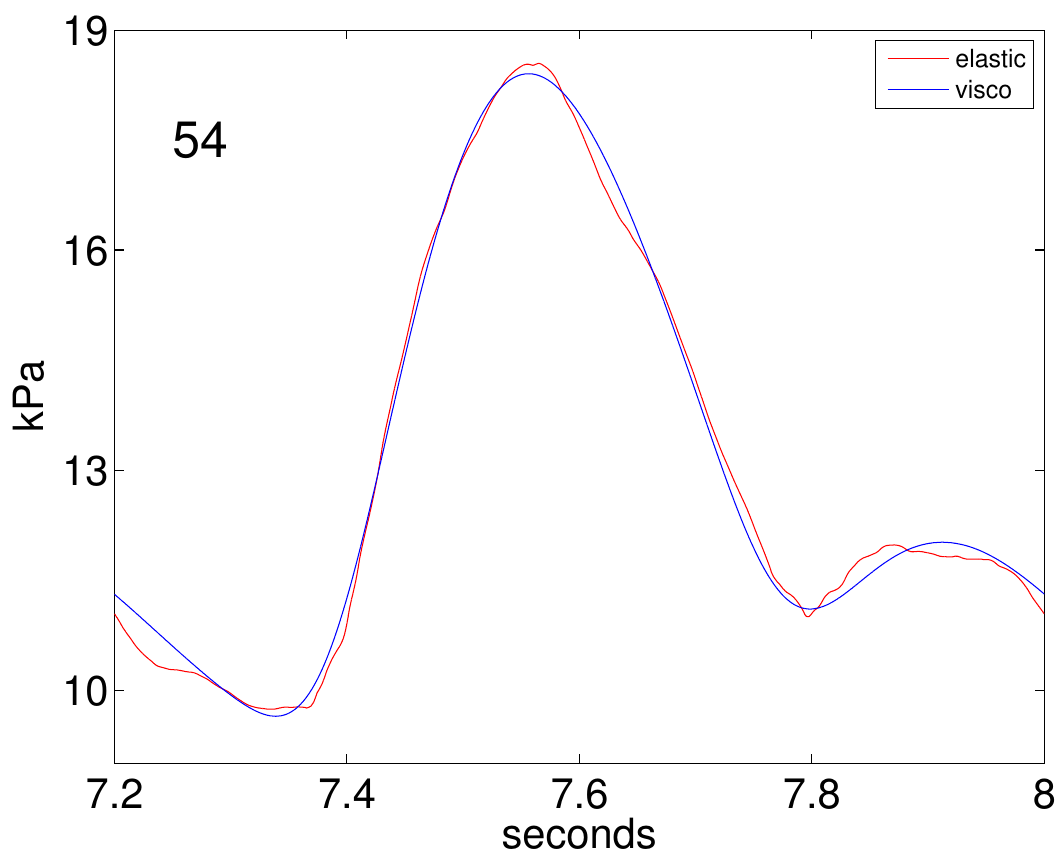} }\\
\caption{The comparison between elastic and viscoelastic models (MUSCL scheme).
The viscoelasticity damps the oscillations of high frequency.}
\label{system_visco}
\end{figure}

Both $\textit{in vivo}$~\cite{reymond2009validation,holenstein1980viscoelastic} and $\textit{in vitro}$~\cite{alastruey2011pulse} 
studies show that the models with viscoelasticity predict the pulse waves better.  
This effect is most pronounced at the peripheral sites~\cite{segers1997assessment,alastruey2011pulse}.
The predictions by the elastic and viscoelastic models are compared at two locations, see Figure~\ref{system_visco}.
We can clearly see the smoothing effect on the pulse curves at both sites. 
The biggest relative difference is observed on the flow rate curve at the peripheral site (see Figure~\ref{fig:Flux_elas_visco54}).  
This study confirms again the necessity to consider the viscoelasticity in the 1D model.

\section{Conclusions} 
In this paper, we incorporated a Kelvin-Voigt viscoelastic constitutive relation of arterial wall with a 1D blood flow model.
This led to a hyperbolic-parabolic system which was then solved by four numerical schemes: MacCormack, Taylor-Galerkin,
MUSCL and local discontinuous Galerkin. The implementations were verified with analytical, semi-analytical or 
clinical observations in many cases.
At first, a single uniform tube was considered.
Under the assumption of small nonlinearities, we obtained asymptotic solutions of the linearized system with different source terms.
The propagation, attenuation and diffusion of the waveform were illustrated by both the numerical and analytical solutions. 
Moreover, in case of a larger nonlinearity, the shock capturing property of each scheme was tested.
After the test on a single vessel, a simple bifurcation was computed to check the numerical coupling of different arteries. 
Finally, we computed a relatively realistic network with 55 arteries.
The check of the numerical solutions in all cases was very favorable for all of the schemes.
We can compare the schemes in four aspects: accuracy, shock-capturing property, computational speed and implementation complexity.

\begin{enumerate}
\item MacCormack and Taylor-Galerkin schemes generate small oscillations. 
MUSCL scheme has slight arbitrary steepening effect. Both diffusion and 
dispersion errors are very small for DG.  
Nevertheless all of the schemes converge with a moderate fine mesh and precisely capture
the various phenomena of this hyperbolicity-dominated hyperbolic-parabolic system.
\item MacCormack, Taylor-Galerkin and DG generate spurious oscillations when the solution is near a shock. 
Numerical flux limiters are possible to filter out the oscillations. 
That will further complicate the schemes and both the theory and technique are still under research~\cite{kuzmin2012slope,marchandise2010accurate}.
On the other hand, there are very mature techniques to impose a slope limiter in the FV scheme.
Shock capturing property is unique for MUSCL among the four schemes presented in this paper. 
But it is very diffusive at a shock, thus a very fine mesh is necessary when a shock may appear.
 
\item For a network of human size, the speed of computation can be ordered from fast to slow as:
Taylor-Galerkin, MUSCL, MacCormack and local DG.
The temporal integration in the Taylor-Galerkin scheme is more efficient than
Adams-Bashforth 2-step method.
Thus it allows a larger time step with a comparable accuracy. 
But if the number of elements for one artery is too large (larger than 500),
Taylor-Galerkin becomes slower because the sizes of the global 
matrices increase quadratically and thus the storing and inverting of matrices become very expensive.
The DG scheme prevents the application of Crank-Nicolson method on the diffusive term.
An explicit method called local DG scheme was adopted in this paper. 
Even with a moderate diffusion coefficient (within the range observed in physiological condition), 
a very small time step is necessary for stability. 
To compute one heart beat, the local DG takes about 9 hours while all other schemes take only 20-90 minutes (using one processor on a standard Linux workstation with MATLAB).

\item From easiest to hardest, the implementation of the schemes can be ordered: MacCormack, MUSCL, Taylor-Galerkin 
and local DG.
\end{enumerate}

As a final conclusion from the point of view of practical application, we recommend MacCormack in case of small nonlinearities as it is very simple and robust.
MUSCL will be a very good option if there may be shock-like phenomena in the system.
Taylor-Galerkin has quite balanced properties between speed and accuracy if no shock-like phenomena may present in the system. 
Local DG is suitable for systems with very small physical diffusive terms since both the numerical diffusion and dispersion are very small in this scheme. 

\section*{Acknowledgements}
This work was supported by French state funds managed by CALSIMLAB and the ANR within the investissements d'Avenir programme under reference ANR-11-IDEX-0004-02. 
The first author also would like to thank the partial financial aid of China Scholarship Council.
We wish to gratefully thank Jean-Fr\'ed\'eric Gerbeau (INRIA) for helpful discussion and implementation of the Taylor-Galerkin scheme, 
and Olivier Delestre (Universit\'e de Nice Sophia-Antipolis) for finite volume scheme.
We are also very grateful to the anonymous reviewers, whose comments helped us a lot to improve this paper.

\bibliographystyle{plain} \bibliography{Numerical}

\end{document}